\documentclass[numerical,a4paper,nofootinbib]{article}
\pdfoutput=1
\usepackage[top=3.3cm, bottom=3.7cm, left=2.5cm, right=3cm]{geometry}
\usepackage{graphicx}
\usepackage{calc}
\usepackage{amsmath}
\usepackage{bbm}
\usepackage{amsfonts}
\usepackage{amssymb}
\usepackage{amsbsy}
\usepackage[T1]{fontenc}
\usepackage{theorem}
\usepackage{stmaryrd}
\usepackage{tipa}
\usepackage{textcomp}
\usepackage{subfigure}
\usepackage{epsfig}
\usepackage{lmodern}
\usepackage{framed}
\usepackage[subfigure]{tocloft}
\usepackage{subeqnarray}
\usepackage{alphalph}
\usepackage[refpage,cfg,prefix]{nomencl}
\usepackage{multirow}
\usepackage{xifthen}
\usepackage{enumerate}
\usepackage{color}
\usepackage[bottom]{footmisc}
\usepackage{natbib}  
\usepackage{wasysym}
\usepackage{bibentry}
\usepackage{verbatim}
\usepackage{todo}
\usepackage{url}
\usepackage{setspace}
\usepackage{xr}
\usepackage{prettyref}

\usepackage{color,colortbl}
\usepackage{relsize}

\makenomenclature

\newcommand{\D}[2]{\frac{\partial #1}{\partial #2}}
\newcommand{\DD}[2]{\frac{\partial^2 #1}{\partial #2^2}}

\newcommand{\ud}{\mbox{d}}

\newcommand{\N}{\ensuremath{\mathbb{N}}}

\newcommand{\listofalgorithms}{\textbf{\Huge{List of Algorithms}}}

\newlistof{Algorithm}{exp}{\listofalgorithms}

\newif\ifletter
\def\ack{\ifletter\bigskip\noindent\ignorespaces\else
    \section*{Acknowledgments}\fi}

\definecolor{light-gray}{gray}{0.4}
\usepackage{epstopdf}

\begin{document}

\title{\textbf{Modelling persistence of motion in a crowded environment: the diffusive limit of excluding velocity-jump processes}}
\author{\textbf{Enrico Gavagnin\footnote{Corresponding author: e.gavagnin@bath.ac.uk}   \,and Christian A. Yates} \\ \small{\textit{Department of Mathematical Sciences},}\\ \small{\textit{University of Bath, Claverton Down, Bath, BA2 7AY, UK}}}

\date{}
\maketitle

\begin{abstract}

Persistence of motion is the tendency of an object to maintain motion in a direction for short time scales without necessarily being biased in any direction in the long term. One of the most appropriate mathematical tools to study this behaviour is an agent-based  \textit{velocity-jump process}. In the absence of agent-agent interaction, the mean-field continuum limit of the agent-based model (ABM) gives rise to the well known hyperbolic telegraph equation. When agent-agent interaction is included in the ABM, a strictly advective system of partial differential equations (PDEs) can be derived at the population-level. However, no diffusive limit of the ABM has been obtained from such a model. Connecting the microscopic behaviour of the ABM to a diffusive macroscopic description is desirable, since it allows the exploration of a wider range of scenarios and establishes a direct connection with commonly used statistical tools of movement analysis.

In order to connect the ABM at the population-level to a diffusive PDE at the population-level, we consider a generalisation of the agent-based velocity-jump process on a two-dimensional lattice with three forms of agent interaction. This generalisation allows us to take a diffusive limit and obtain a faithful population-level description. We investigate the properties of the model at both the individual and population-level and we elucidate some of the models' key characteristic features. In particular, we show an intrinsic anisotropy inherent to the models and we find evidence of a spontaneous form of aggregation at both the micro- and macro-scales. \\

    
\end{abstract}

\maketitle
\section{Introduction}
\label{sec:introduction}
Understanding the properties of cell movement is of fundamental interest in many biological contexts such as embryogenesis \citep{mort2016rdm}, epidermal wound healing \citep{maini2004twm} and tumour growth \citep{sherratt2001nmm}. 
Mathematical models are now considered essential tools in cell biology for testing theoretical hypotheses, interpreting experimental data and extracting biological parameters \citep{simpson2007sic,simpson2009mss,johnston2014isa,ross2017uab}. 
There are typically two approaches to modelling cell motion, either micro-scale discrete \citep{anderson1998cdm,simpson2007sic, peirce2004msp,deutsch2007cam, alber2003caa, segovia2004icm, wang2015scg} or macro-scale continuum \citep{simpson2006lii,murray2007mbi,anderson1998cdm,wise2008tdm}. 
The discrete approach, using agent-based models (ABMs), accounts for properties at the cell-scale, while the continuum approach, often presented as a system of partial differential equations (PDEs) or stochastic partial differential equations (SPDEs), gives a global description of the migration at the population-level. Continuum models have the advantage that they are generally more amenable to mathematical analysis and can lead to significant insights for situations in which the system comprises a large number of agents, at which point simulating the ABM becomes computationally expensive. Nevertheless, finding the appropriate continuum model to describe the collective behaviour of a system of moving agents can be a difficult task and continuum models are often specified on a phenomenological basis, which may reduce their predictive power. It is essential, therefore, to establish a connection between micro-scale properties, which can be inferred directly from experimental data, and macro-scale dynamics \citep{noble2002orc,baker2010fmm,yates2012gfm}. 

Many analyses of cell migration are based on the hypothesis that the movement of a single cell can be described as a simple random walk  on a lattice \cite{simpson2009mss, mort2016rdm, baker2010fmm, deutsch2007cam}. In many models, the behaviour of a single cell is assumed to be independent of the other cells' positions and multiple cells can occupy the same lattice site simultaneously \cite{codling2008rwm, berg1972cec}. In many applications, however, crowding effects play an important role that can not be neglected \cite{simpson2007sic,maini2004twm}. Crowding is incorporated into such models via volume exclusion: each lattice site is allowed to be occupied by at most one cell \cite{treloar2011vjm,treloar2012vjp,slowman2016jai,othmer1997abc,stevens2000sca}. A macroscopic continuum description of this type of model can be obtained by considering an average mass conservation law for each lattice site and taking an appropriate limit as the spatial and temporal discretisation steps go to zero simultaneously \cite{deutsch2007cam}. 

One of the key aspects of a simple random walk is that the direction of motion undergoes a series of uncorrelated jumps in space. In reality, experimental observations indicate that many types of cell tend to preserve their direction of motion for a certain time before reorienting \cite{berg1972cec, hall1977amc, gail1970lmf, wright2008def}, even though the movement is globally unbiased. This tendency is normally called \textit{persistence of motion} or \textit{of direction} \cite{patlak1953rwp}. There is vast literature about modelling persistence at multiple-scales for non-interacting agents \cite{othmer1988mdb,othmer2000dlt, othmer2002dlt2,codling2008rwm,campos2010prm}, but it is only in recent years that there has been an increasing interest in studying the role of persistence of motion for systems of self-interacting agents  \cite{treloar2011vjm,treloar2012vjp,slowman2016jai,thompson2011lmn,sepulveda2016ccr,soto2014rtd,peruani2006ncs,redner2013sdp} and persistence induced by crowding \cite{grima2010cia,smith2017mcd}.

Typically, to incorporate persistence in the ABM, cell movement is represented as a correlated random walk (CRW) which is also known as a \textit{velocity jump process} \cite{codling2008rwm,othmer1988mdb}. In this model, the cell has an assigned direction of motion (left or right in one dimension) and it moves in this direction with constant velocity, $v$, until the assigned direction is changed, which occurs according to a Poisson process with a given rate, $\lambda$. Notice that the temporary preferential direction induces bias in the motion for short time-scales, which represents persistence, but the resulting motion remains globally unbiased for longer time scales. 
  
In the case of non-interacting agents, the macro-scale behaviour of the velocity-jump process in one dimension is well known to evolve according to the hyperbolic telegraph equation for the cell density $C(x,t)$ \cite{kac1974smr, goldstein1951ddm, othmer1988mdb}:
 \begin{equation}
 	\label{eq:telegraph}
 	\DD{C}{t}+2\lambda \D{C}{t}=v^2 \DD{C}{x} \; .
 \end{equation}
Notice that Eq.\,\eqref{eq:telegraph} was originally developed to describe the propagation of waves which travel and reflect trough a telegraph transmission line \citep{metzger2012tlp}. The same type of equation can be derived from a system of non-interacting agents performing a velocity-jump process in one dimension. In particular,  \citet{othmer1988mdb} derive the telegraph equation for cells undergoing velocity-jump processes without interactions.  \citet{othmer2000dlt} demonstrated that it is possible to obtain a parabolic limit as $v$ and $\lambda$ tend to infinity simultaneously, such that  $v^2/\lambda$ remains constant. In this limit the canonical diffusion equation is recovered \cite{othmer2000dlt,othmer2002dlt2}. This is not a surprise, since the short-term correlation effects become less evident at large time scales and so the limit process is effectively equivalent to a simple random walk. 
 
 When direct agent-agent interactions are introduced, however, the derivation of an exact closed form PDE for the total mean agent density is not possible \citep{codling2008rwm}.  Recently, \citeauthor{treloar2011vjm} have derived a system of macroscopic advective equations from a velocity-jump process with three different forms of direct interaction \cite{treloar2011vjm,treloar2012vjp}. Although their continuum model is successful in replicating the population-level behaviour of the ABM for a limited range of model parameters, the first order approximation considered in \citet{treloar2011vjm} enforces restrictions in the initial condition (which must be sufficiently smooth) and on the choice of the parameters. 
  
%------------------------
The aim of our work is to ease these restrictions in order to provide a better connection between discrete and continuum  models of volume excluding persistent agents. We consider a generalisation of the ABM of \citet{treloar2011vjm} in which we modulate the influence of persistence through an additional parameter $\varphi$. This allows us to take a \textit{diffusive} limit if the new parameter $\varphi$ scales with the lattice size. The resulting PDE description includes a non-linear diffusive term, which encapsulates the long-term diffusive behaviour of cells, and an advective part, which is consistent with the findings of \citet{treloar2011vjm}. Our new diffusive model represents an extension of the previous advective model and can be applied to study a wider range of scenarios. In particular, we can consider situations with steep gradient in cell density, which have not previously been investigated. Moreover, a diffusive limit is appropriate for the study of the long term behaviour of the system, especially if we are interested in statistical tools which are related to the diffusion coefficient, such as the mean squared displacement and the mean dispersal distance \cite{codling2008rwm}.

%------------------------

In this paper we study the two-dimensional version of a model which incorporates persistence and volume exclusion. We explain the derivation of the diffusive continuum description and finally we test the agreement between our discrete and continuum models using some illustrative examples. Our new diffusive PDEs correctly represent the population-level behaviour of our ABMs, in particular in scenarios that could not be investigated with the previous advective models. Our investigation highlights some peculiar aspects of the excluding velocity-jump processes, which we discuss in the light of the new macroscopic description. In particular, a spontaneous form of aggregation, similar to that observed by \citet{thompson2011lmn} and \citet{sepulveda2016ccr}, 
appears in both our agent- and population-level models. We believe this is the first reported example in which such a phenomenon appears at both micro- and macro-scales.

The paper is organised as follows. In Section \ref{sec:ABM} we define the ABM and introduce three forms of cell-cell interaction. In Section \ref{sec:cont_approx}, we derive the continuum diffusive description of the ABM from the occupancy master equations. Our numerical results on the comparison between the ABMs and the corresponding PDEs are shown in Section \ref{sec:results}, together with our observations on some of the  interesting model behaviours. We conclude with a short discussion of our results and possible avenues for future research in Section \ref{sec:conclusion}.

\section{The Agent-Based Model}
\label{sec:ABM}
In this section we describe the basic ABM. The models presented in the following sections are all adaptations of this basic model. Cells are represented by agents on a square lattice of size $L_x \times L_y$ sites and lattice step $\Delta$ with periodic boundary conditions in the $y$-direction and zero-flux boundary conditions in the $x$-direction\footnote{We implement periodic boundary conditions in the vertical direction to avoid edge-effects. In the horizontal direction we employ zero-flux boundary conditions, although in reality agents rarely, if ever, reach these boundaries. So other boundary conditions may be employed with little consequence.}. Each site of the lattice can be occupied by at most one cell, in which case we will say the site is occupied, otherwise the site is said to be empty.

\begin{figure}[h!!!!!!!!!!!!!!!!]
\begin{center} 
 \includegraphics[width=0.8\columnwidth]{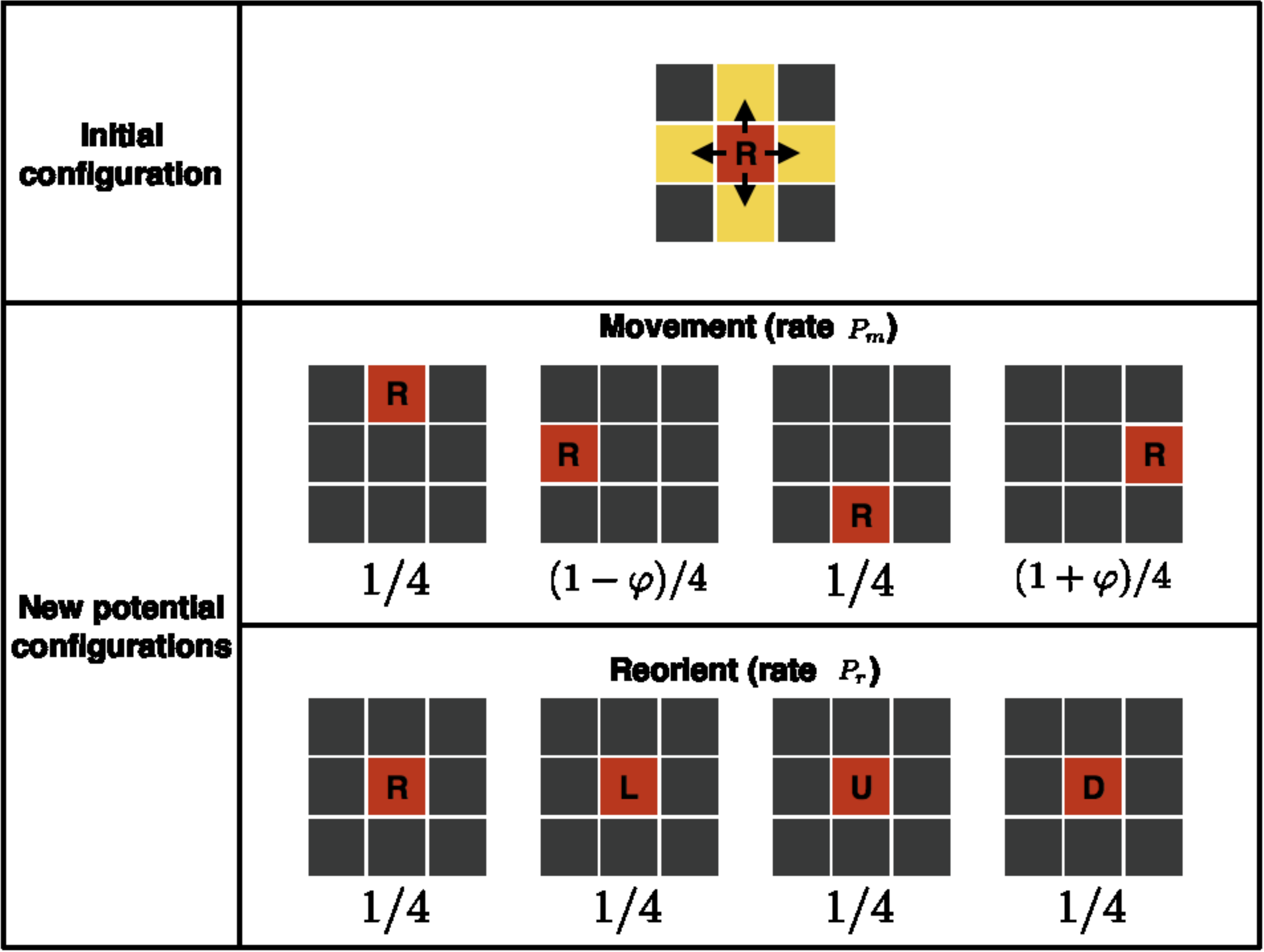}
\end{center}
\caption{Diagram of the motility mechanism of an R-polarised agent in the ABM model with velocity $v=1$. Red (dark grey) sites are occupied by the moving agent with the polarisation denoted by the corresponding letter (R: Right, L: Left, U: Up, D: Down), black sites are empty and yellow (light grey) sites highlight the von Neumann neighbour sites. The top panel shows the initial configuration and the neighbour sites reachable by the agents. The middle panel shows the four new potential configurations in the case that a movement occurs (with rate $P_m$) with the corresponding probabilities. The bottom panel shows the four new potential configurations in the case that a reorientation event occurs (with rate $P_r$), the agent remains in the same site and its polarisation is chosen uniformly at random.}
\label{fig:schematic_multipol}
\end{figure}

We assign to each agent a polarisation in one of the four directions of the lattice. We denote such polarisation with the corresponding initial capital letter: Right (R), Left (L), Up (U) and Down (D). Let $v\in \N^+$ be a positive integer which denotes the number of lattice sites that an agent can move during a single movement event.  We can interpret this as a non-dimensional measure of the agent's velocity. Agents can move or reorient their polarisation in continuous time. Both of these events occur at random as independent Poisson processes with rates $P_m$ and $P_r$, respectively. The role of the polarisation is to induce a temporary bias in the stochastic motion so that the polarised agent is more likely to move in the corresponding direction. Let $\varphi\in[0,1]$ be a parameter which characterises the intensity of the bias. Consider, for example, an R-polarised agent in two dimensions, located at site $(i,j)$. If the agent is chosen to move, one of the four sites $(i\pm v,j), \, (i,j\pm v)$ is selected, at random, as a target site. The right-hand site $(i+v,j)$, corresponding to the R-polarisation of that cell, is chosen with probability given by $\frac{1+\varphi}{4}$. In the opposite direction to the polarisation, the left-hand site $(i-v,j)$ is chosen with probability $\frac{1-\varphi}{4}$ and each of the sites in the vertical direction $(i,j\pm v)$ (orthogonal to the polarisation direction) are chosen with probability $\frac{1}{4}$ (see the schematic in Fig. \hspace{-0.3em }\ref{fig:schematic_multipol})\footnote{Alternative transition probabilities could be considered. For example one could choose to reduce the movement probabilities in the direction orthogonal to the polarisation (Up and Down in the example of an R-polarised agent) as well as in the direction opposite to the current polarisation. A similar derivation can be applied, which will lead to a similar but slightly altered macroscopic model.}. The transition probabilities for the other polarisations are obtained analogously. Finally, if a reorientation event occurs, with rate $P_r$, the agent changes its polarisation uniformly at random to one of the four possible polarisations (including the possibility of maintaining its current polarisation), see  Fig. \hspace{-0.3em }\ref{fig:schematic_multipol}. 

Notice that if $\varphi=0$, then the target site is chosen uniformly and the movement corresponds to a classic uncorrelated random walk (with non-local jumps for $v>1)$. On the contrary, if we let $\varphi=1$ we achieve the strongest bias where agents cannot move in the opposite direction to their polarisation\footnote{Note that if the model is specified in one dimension and $\varphi=1$, the model corresponds to the velocity-jump process described in \citet{treloar2011vjm}. However, in the two-dimensional case the choice of maximum bias leads to a different model. In particular, the target site is still chosen at random (although not uniformly) between three of the nearest neighbours while in \citet{treloar2011vjm}, upon an agent being selected to move, its target site is chosen deterministically.}. 

\begin{figure}[h!!!!!!!!!!!!!!!!]
\begin{center} 
 \subfigure[][]{\includegraphics[width=0.45\columnwidth]{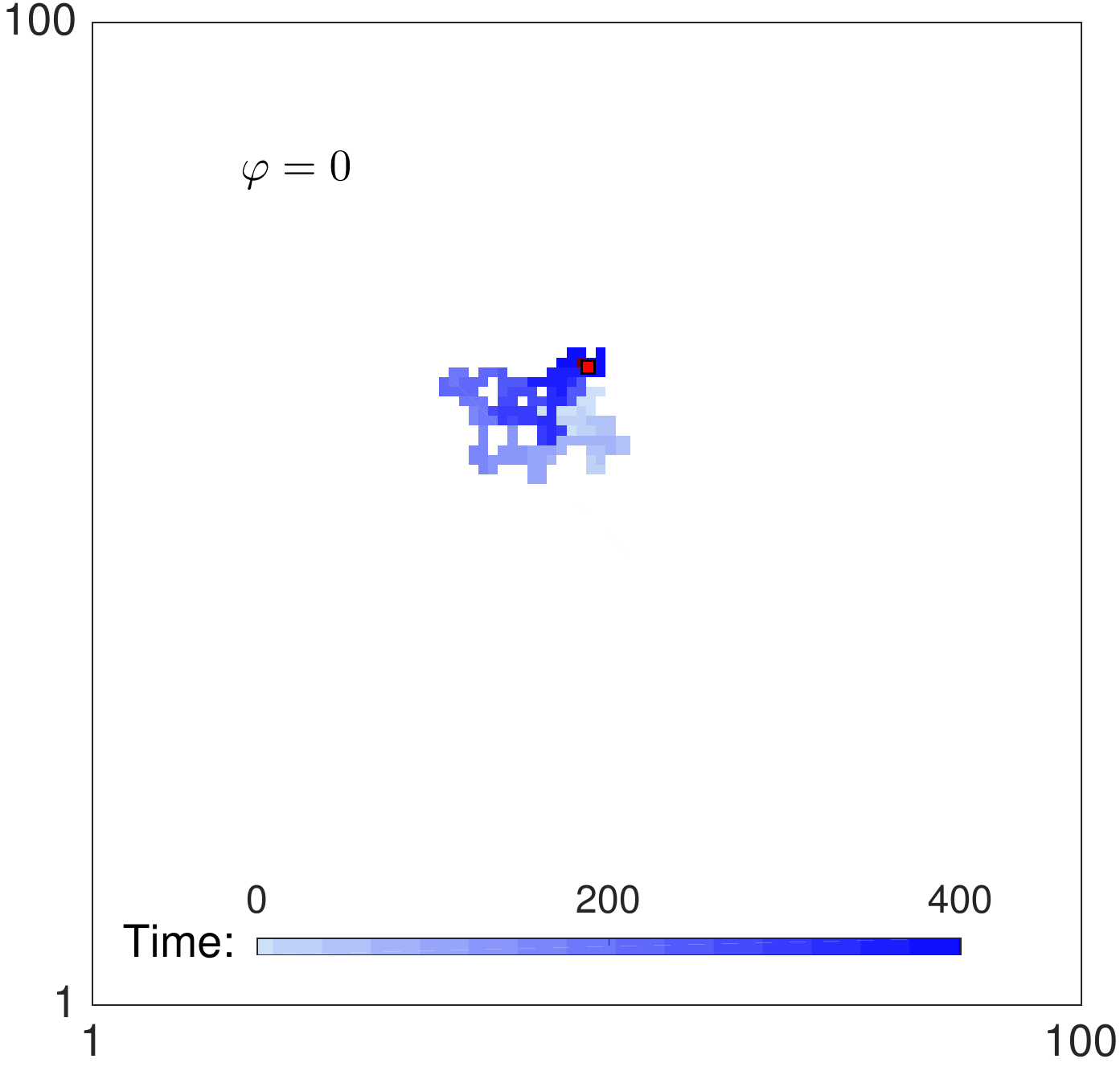}
\label{fig:non-persistent_motion_example}
}
 \subfigure[][]{\includegraphics[width=0.45\columnwidth]{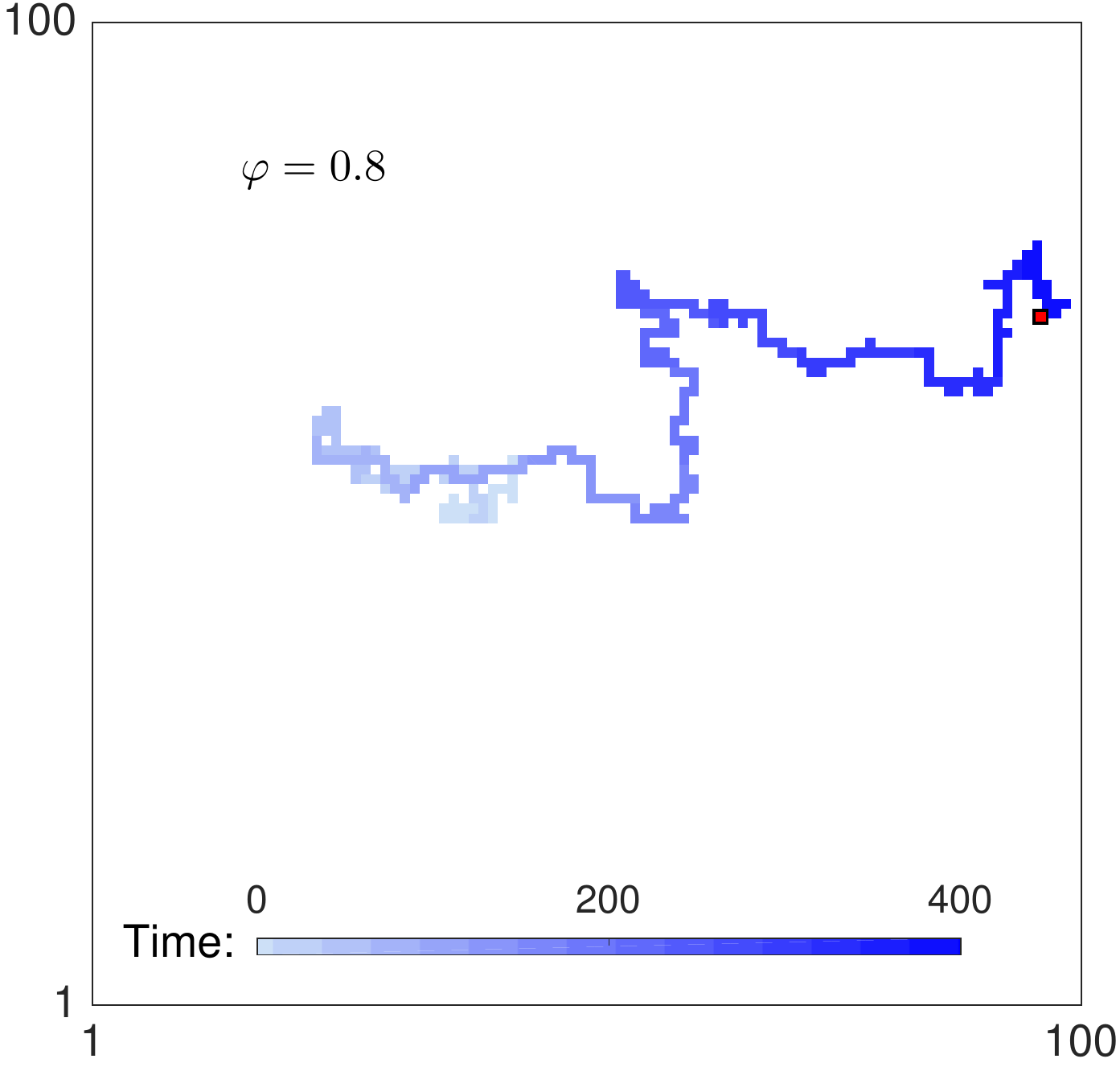}
\label{fig:persistent_motion_example}
}
\end{center}
\caption{Trajectories of a single agent moving according to the ABM scheme on a two-dimensional square lattice with $L_x=L_y=100$ and $\Delta=1$. The left panel (a) shows an example of simple random walk ($\varphi=0$), while the right panel (b) shows an example of persistent random walk ($\varphi=0.8$). The blue (grey) points represent the subsequent positions, from light ($t=0$) to dark ($t=400$) blue (grey), of a single agent initialised in the centre of the domain, $(50,50)$. The last position is highlighted in red (light grey with black border). In both panels the parameters are $P_m=1$, $P_r=0.05$, $v=1$.}
\label{fig:multipol_1}
\end{figure}

When the rates of reorientation and movement are chosen such that $P_r \ll P_m$ together with a large value of the parameter $\varphi$, agents persist in their direction of motion. Fig. \hspace{-0.3em }\ref{fig:multipol_1} shows two trajectories of a single agent for parameters $P_m=1$, $P_r=0.05$, $v=1$ and for panel \subref{fig:non-persistent_motion_example} $\varphi=0$ (no persistence), for panel \subref{fig:persistent_motion_example} $\varphi=0.8$ (strong persistence). Persistence of motion is clearly evident in the shape of the track in panel \subref{fig:persistent_motion_example}. The long term behaviour is unbiased, since none of the four directions is preferred in the long term. As there is only one agent in the domain, exclusion (specified in the next paragraph) does not play any role in the dynamics. 

\vspace{0cm} 

\begin{figure}[h!!!!!!!!!!!!]
\begin{center} 
\vspace{1cm}
 \includegraphics[width=0.55\columnwidth]{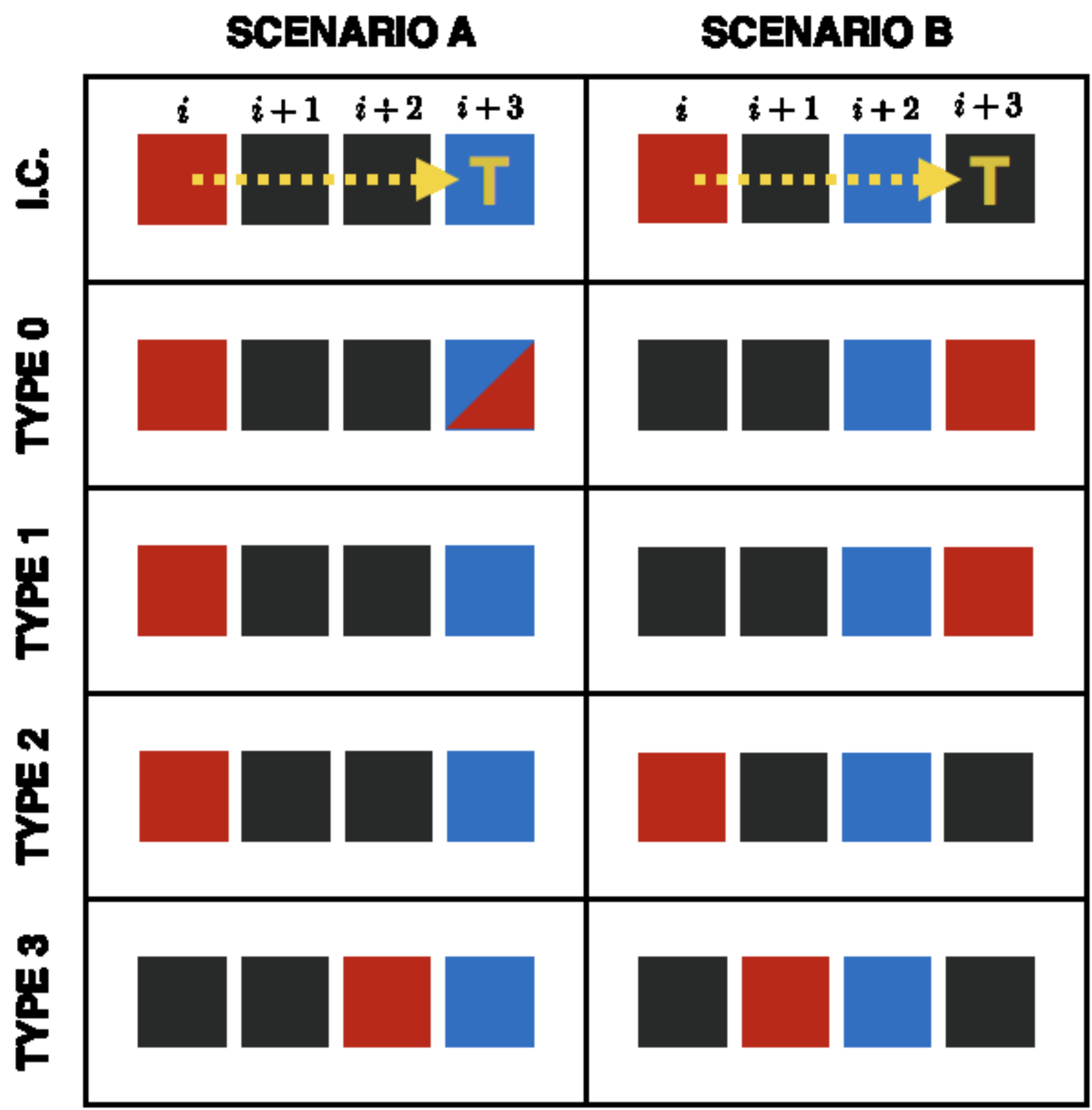}
\end{center}
\caption{Schematics of the four different exclusion properties considered in one dimension. The first row shows the initial configuration for the two scenarios considered. In both cases the moving agent (red, dark grey) attempts to move to the target site (yellow T) three sites to the right ($v=3$). The black sites are empty and the blue (light grey) sites are occupied by other agents. Each of the subsequent four rows represent the new configuration of cells under the exclusion property chosen.}
\label{fig:schematic_exclusion}
\end{figure}

Once the target site is selected, the agent moves according to the exclusion property specified for the process. For consistency with \citet{treloar2011vjm}, we consider four different exclusion properties, one without agent interaction and three with a variety of interactions. Fig. \hspace{-0.3em }\ref{fig:schematic_exclusion} shows two typical scenarios (for $v=3$) in which an agent (red) at position $i$ attempts to move to the target site at $i+3$. In Scenario A the target site is occupied by another agent (blue), while in Scenario B the target site is empty and the site $i+2$ is occupied. We use these two examples to explain the four exclusion properties as follows. 

\paragraph*{Type 0 - Non-interacting agents.} \label{sec:ABM_a} In this case the moving agent moves to the target site regardless of its occupancy. Such a process is not an exclusion process since arbitrarily many agents can occupy the same site. In both scenarios in Fig. \hspace{-0.3em }\ref{fig:schematic_exclusion} the moving agent moves to the target site. In Scenario A the agent shares the site with the other agent (blue) and in Scenario B it occupies the target site alone.

\paragraph*{Type 1 - Only move if target site is available.} \label{sec:ABM_b} In this case the agent moves only when the target site is not occupied. In Scenario A of Fig. \hspace{-0.3em }\ref{fig:schematic_exclusion} this exclusion property causes the entire movement to be aborted, while in Scenario B the moving agent jumps over the blue agent occupying the site $i+2$ in order to reach site $i+3$. 

\paragraph*{Type 2 - Farsighted agents.} \label{sec:ABM_c} In this case the movement takes place only if the target site and all the intermediate sites are vacant. If at least one of the sites is occupied, the movement is aborted and the moving agent remains at the initial position. The assumption distinguishing this exclusion process is that agents know the occupancy of distant sites in order to decide whether to move to the target or not. If cells extend sensing filopodia, this behaviour can be justified for short distances \cite{wolf2007mpp}, but it becomes unrealistic for large values of $v$. In both scenarios of Fig. \hspace{-0.3em }\ref{fig:schematic_exclusion} the movement under this movement type is aborted since either the target or an intermediate site is occupied. 
\paragraph*{Type 3 - Shortsighted agents.} \label{sec:ABM_d} This is the most mathematically sophisticated and realistic form of interaction that we consider. With this scheme, the moving agent moves through the intermediate sites between its position and the target site and stops at the furthest site which it can reach without being blocked by any of the other agents. If no blocking occurs, the agent moves to the target site. In Fig. \hspace{-0.3em }\ref{fig:schematic_exclusion} we see that this exclusion property allows the agent to move in both scenarios, the distance  depending on the position of the blocking agent (blue).

\section{Population-Level Model}
\label{sec:cont_approx}
In this section we  derive a family of PDEs which describe the behaviour of the ABMs introduced in Section \ref{sec:ABM} at the population-level. The general technique consists of writing down the continuous-time occupancy master equation for the average occupancy of a general site of the ABM, Taylor expanding and finally taking a limit as the lattice step, $\Delta$, and the time step, $\tau$, go to zero jointly while holding $\Delta^2/\tau$ constant \cite[Chapter 4]{deutsch2007cam}. The four types of agent interactions considered lead to different PDEs, therefore, in what follows, we will consider those cases in different subsections.

We denote by $C^t_{i,j}(m)$ the occupancy of the site $(i,j)$ at time $t$ for the $m$-th simulation of the ABM, \textit{i.e.} $C^t_{i,j}(m)=1$ if the site $(i,j)$ at time $t$ in the $m$-th simulation is occupied and $C^t_{i,j}(m)=0$ if it is empty. We denote with $\hat{C}^t_{i,j}$ the occupancy of the site $(i,j)$ at time $t$ averaged over all $M$ realisations, \textit{i.e.}:
\begin{equation}
\label{eq:def_average_occupancy}
\hat{C}^t_{i,j}=\frac{1}{M}\sum_{m=1}^M C^t_{i,j}(m)  .
\end{equation}
In addition, we define occupancy variables for the four subpopulations of agents classified according to their polarisation (Right, Left, Up, Down) and we denote these occupancies with the capital corresponding to their first letter. For example, $R^t_{i,j}(m)$ represents the occupancy of right-polarised agents at the site $(i,j)$ at time $t$ for the $m$-th simulation and $\hat{R}^t_{i,j}$ the corresponding  occupancy averaged over the total number of realisations. The total occupancy can be obtained by adding together the occupancies of the four subpopulations as $\hat{C}^t_{i,j}=\hat{R}^t_{i,j}+\hat{L}^t_{i,j}+\hat{U}^t_{i,j}+\hat{D}^t_{i,j}$. From now on, we omit the hats in the averaged occupancies for simplicity. All the occupancies are defined for $t\in \mathbb{R}^+$ and $(i,j) \in \lbrace1, \dots , L_x\rbrace \times \lbrace 1,\dots ,L_y\rbrace$. 

\subsection*{Type 0 interaction}
\label{sec:PDE_a}
For the case of non-interacting agents we can write down the occupancy master equations of the process, in sites away from the boundary\footnote{Note for sites on the boundary, the occupancy equations will be slightly different, but we will employ zero-flux or periodic boundary conditions (respectively) on the vertical and horizontal boundaries (respectively) in the continuum model to match our specification in the ABM.}, as
\begin{equation}
\begin{split}
	\label{eq:master_0}
	%equation for R
	&\begin{aligned}
	R^{t+\tau}_{i,j}=R^t_{i,j}&+\frac{\tau P_m}{4} \left[ (1+\varphi )R^t_{i-v,j}+ (1-\varphi)R^t_{i+v,j} +  R^t_{i,j+v}+R^t_{i,j-v} -4R^t_{i,j} \right] \\
	&+ \frac{\tau P_r}{4} \left[ L^t_{i,j}+U^t_{i,j}+D^t_{i,j}-3R^t_{i,j}\right]+\mathcal{O}(\tau^2),
	\end{aligned} \\[15pt]
	%equation for L
	&\begin{aligned}
	L^{t+\tau}_{i,j}=L^t_{i,j}&+\frac{\tau P_m}{4} \left[ (1-\varphi )L^t_{i-v,j}+ (1+\varphi)L^t_{i+v,j} +  L^t_{i,j+v}+L^t_{i,j-v} -4L^t_{i,j} \right] \\
	&+ \frac{\tau P_r}{4} \left[ R^t_{i,j}+U^t_{i,j}+D^t_{i,j}-3L^t_{i,j}\right]+\mathcal{O}(\tau^2),
	\end{aligned}\\[15pt]
	%equation for U
	&\begin{aligned}
	U^{t+\tau}_{i,j}=U^t_{i,j}&+\frac{\tau P_m}{4} \left[ U^t_{i+v,j}+U^t_{i-v,j}+(1-\varphi )U^t_{i,j+v}+ (1+\varphi)U^t_{i,j-v} -4U^t_{i,j} \right] \\
	&+ \frac{\tau P_r}{4} \left[ R^t_{i,j}+L^t_{i,j}+D^t_{i,j}-3U^t_{i,j}\right]+\mathcal{O}(\tau^2),
	\end{aligned}\\[15pt]
	%equation for D
	&\begin{aligned}
	D^{t+\tau}_{i,j}=D^t_{i,j}&+\frac{\tau P_m}{4} \left[ D^t_{i+v,j}+D^t_{i-v,j}+(1+\varphi )D^t_{i,j+v}+ (1-\varphi)D^t_{i,j-v} -4D^t_{i,j} \right] \\
	&+ \frac{\tau P_r}{4} \left[ R^t_{i,j}+L^t_{i,j}+U^t_{i,j}-3D^t_{i,j}\right]+\mathcal{O}(\tau^2) ,
	\end{aligned}
	\end{split}
\end{equation}
where $\tau$ is sufficiently small such that the probability that more than one event occurs in the time interval $(t,t+\tau)$ is  $\mathcal{O}(\tau^2)$.
The terms in system \eqref{eq:master_0} that contain $P_m$ represent the transitions into and out of the site $(i,j)$ due to the motility events, while the terms which contain $P_r$ represent the transition from one polarisation to another due to the reorienting events. In order to obtain a continuous approximation of the model, we Taylor expand terms such as $R^t_{i\pm v,j}$ and $R^t_{i,j\pm v}$ about position $(i,j)$ as:
\begin{equation}
\label{eq:taylor}
\begin{split}
	&R^t_{i\pm v,j}=R(x_{i\pm v},y_j,t)=R(x_i,y_j,t) \pm v \Delta \D{R}{x} (x_i,y_j,t)+\frac{1}{2} (v \Delta)^2  \DD{R}{x} (x_i,y_j,t)+ \dots ,	\\
	&R^t_{i,j\pm v}=R(x_{i},y_{j\pm v},t)=R(x_i,y_j,t) \pm v \Delta \D{R}{y} (x_i,y_j,t)+\frac{1}{2} (v \Delta)^2  \DD{R}{y} (x_i,y_j,t)+ \dots .	
\end{split}
\end{equation}
In order to take the diffusive limit we assume that $\varphi$ rescales with the spatial step of the lattice, \textit{i.e.} $\varphi=\mathcal{O}(\Delta)$. This assumption, that the local bias tends to zero, is necessary in order to derive a finite advective term in the diffusive limit. A similar assumption is made in order to derive consistent continuum limits of models with global bias \cite{codling2008rwm,simpson2009mss}. 

By substituting the truncated Taylor expansions (of which equations \eqref{eq:taylor} are an example) into equations \eqref{eq:master_0} and rearranging terms, we can take the diffusive limit as $\tau, \Delta \rightarrow 0$ in equations \eqref{eq:master_0} such that $\Delta^2/\tau$ remains fixed. We obtain a system of PDEs for the continuous occupancy functions $R(x,y,t),L(x,y,t),U(x,y,t),D(x,y,t)$, where $t\in \mathbb{R}^+$ and $(x,y)\in[0,\Delta L_x]\times [0,\Delta L_y]$,

\begin{equation}
\begin{split}
	\label{eq:PDE_0}
	%equation for R
	&\begin{aligned}
	\D{R}{t}=v^2 \mu \nabla^2 R - v\nu \D{R}{x}+\frac{P_r}{4}(C -4 R) 
	,\end{aligned} \\
	%equation for L
	&\begin{aligned}
	\D{L}{t}=v^2 \mu \nabla^2 L+v \nu \D{L}{x}+\frac{P_r}{4}(C -4 L) 
	,\end{aligned} \\
	%equation for U
	&\begin{aligned}
	\D{U}{t}=v^2\mu\nabla^2 U - v\nu \D{U}{y}+\frac{P_r}{4}(C -4 U) 
	,\end{aligned} \\
	%equation for D
	&\begin{aligned}
	\D{D}{t}=v^2\mu\nabla^2 D +v \nu \D{D}{y}+\frac{P_r}{4}(C -4 D) 
	,\end{aligned} \\
	\end{split}
\end{equation}
where 
\begin{equation}
\label{eq:coeff}
	\mu:=\lim_{\Delta \rightarrow 0} 	\frac{ \Delta^2 P_m}{4}  \qquad \text{and} \qquad \nu:=\lim_{\Delta \rightarrow 0} 	\frac{\varphi \Delta P_m}{2}  .
\end{equation}
The boundary conditions are chosen to be consistent with the ABM. In particular, for every $x\in [0,\Delta L_x]$, $y \in [0, \Delta L_y]$ and $t\in \mathbb{R}^+$ we impose
\begin{equation}
\label{eq:BC_PDE}	
\begin{split}
R(x,0,t)=R(x,\Delta L_y,t) , &\qquad \D{R}{x}(0,y,t)=\D{R}{x}(\Delta L_x ,y,t)=0 \, ,\\
L(x,0,t)=L(x,\Delta L_y,t) , &\qquad \D{L}{x}(0,y,t)=\D{L}{x}(\Delta L_x ,y,t)=0 \,,\\
U(x,0,t)=U(x,\Delta L_y,t), &\qquad \D{U}{x}(0,y,t)=\D{U}{x}(\Delta L_x ,y,t)=0 \,, \\
D(x,0,t)=D(x,\Delta L_y,t), &\qquad \D{D}{x}(0,y,t)=\D{D}{x}(\Delta L_x ,y,t)=0 \,.
\end{split}
\end{equation}
The two limits in \eqref{eq:coeff} exist and are finite owing to the assumption on $\varphi$ above. The right-hand sides of system \eqref{eq:PDE_0} comprise three terms (in order from left to right): a diffusive term, an advective term and a reactive term. The diffusive terms capture the long-term unbiased motion of the agents. In the case of the non-interacting agents, described above, the diffusion coefficient is independent of the agent density. The advective terms reflect the polarisation of each subpopulation and, as such, they involve the first partial derivative of density in the direction of the polarisation. The reactive terms represent the uniform changing of polarisation. 

We can write down the PDE for the total averaged density by adding the equations of the system \eqref{eq:PDE_0}:
\begin{equation}
\begin{aligned}
	\label{eq:tot_PDE0}
	\D{C}{t}=v^2\mu \nabla^2 C + v\nu \D{}{y} \left[(D-U)\right]+ v\nu \D{}{x} \left[ (L-R)\right]  \; .
	\end{aligned}
	\end{equation}
Note that a closed form for the equations in terms of the total density is not possible unless $\varphi=0$, in which case the total density evolves according to the canonical diffusion equation, consistently with \cite{othmer1988mdb,codling2008rwm}. 

Notice that advective-diffusive equations like the one of system \eqref{eq:PDE_0} are an extensively studied class of PDEs which can be found in a wide range of applications. In particular, they are traditionally used to represent transport phenomena such as heat transfer \citep{bird2002tph,welty2009fmh}, mass transfer \citep{welty2009fmh} and virus propagation \citep{tim1991mpv,sim2000vtu}.

\subsection*{Type 1 interaction}
\label{sec:PDE_b}

For the first non-trivial type of interaction that we consider, the movement of the agents depends only on the occupancy of the target site. Specifically, movement is aborted if, and only if, the target site is occupied (see Fig. \hspace{-0.3em }\ref{fig:schematic_exclusion}). The occupancy master equation for the right-subpopulation reads:
\begin{align}
\label{eq:ME-1}
\begin{split}
	R^{t+\tau}_{i,j}&=R^t_{i,j}\\&+\frac{\tau P_m}{4} \left(1-C^t_{i,j}\right) \left[ (1+\varphi )R^t_{i-v,j}+ (1-\varphi)R^t_{i+v,j} +  R^t_{i,j+v}+R^t_{i,j-v} \right] \\
	& -\frac{\tau P_m}{4} R^t_{i,j} \left[ (1+\varphi ) \left(1-C^t_{i+v,j}\right)+ (1-\varphi)\left(1-C^t_{i-v,j}\right) +  \left(1-C^t_{i,j+v}\right) +\left(1-C^t_{i,j-v}\right)\right]\\
	&+ \frac{\tau P_r}{4} \left[ L^t_{i,j}+U^t_{i,j}+D^t_{i,j}-3R^t_{i,j}\right] +\mathcal{O}(\tau^2).
	\end{split}
\end{align}
The equations for the other three subpopulations are given in Appendix \ref{sec:OME}. The main difference in comparison to the non-interacting type 0 models is the introduction of terms that reduce the probability of moving according to the density of the target site. For example, the term $(1-C^t_{i,j})$ determines probability of success of a movement into the site $(i,j)$ at time $t$. If at time $t$ the site $(i,j)$ is occupied in all the realisations of the ABM we have $C^t_{i,j}=1$ so the corresponding probability of success is zero. \textit{Vice versa} if the site $(i,j)$ is empty in all the simulations, the probability of success is one, since the movement is always allowed to take place. Notice that in writing down the occupancy master equation \eqref{eq:ME-1}, we are making the mean-field assumption that the occupancies of neighbouring sites are independent.

We can use the same steps as in Section \ref{sec:PDE_a}, for type 0 interactions, to obtain a system of diffusive PDEs for the density of the four different polarisations. The resulting equation for the right-moving subpopulation is given by 
\begin{equation}
\begin{split}
	\label{eq:PDE_1}
	%equation for R
	&\begin{aligned}
	\D{R}{t}=v^2\mu \left[R \nabla^2 C+(1-C) \nabla^2 R\right] - v\nu \D{}{x} \left[ R(1-C)\right]+\frac{P_r}{4}(C -4 R) 
	,\end{aligned} \\
	\end{split}
\end{equation}
where $\mu$ and $\nu$ are as defined in equations \eqref{eq:coeff}. See system \eqref{eq:PDE_1} of the Appendix \ref{sec:CDS} for the full set of equations. We can see that the switching rates between subpopulations remain the same as in system \eqref{eq:PDE_0}, whereas the advective and the diffusive terms of equation \eqref{eq:PDE_1} depend linearly on the cell density. Specifically, both the advective and the original diffusive parts are scaled by a factor $1-C$, which takes into account the decrease in motility due to the volume exclusion. An additional diffusive term $\nabla^2 C$ appears, scaled by the density of each subpopulation. Notice that, by adding together the equations for the four subpopulations, we recover a normal diffusive term for the total cell density:
\begin{equation}
	\begin{aligned}
	\label{eq:tot_PDE1}
	\D{C}{t}=v^2\mu \nabla^2 C + v\nu \D{}{y} \left[ (D-U)(1-C)\right]+ v\nu \D{}{x} \left[ (L-R)(1-C)\right]  \; .
	\end{aligned}
\end{equation} 
Nevertheless, as in the previous case (type 0), the advective terms make it impossible to close the PDE for the total density, $C(x,y,t)$, apart from in the trivial case $\varphi=0$.

\subsection*{Type 2 interaction}
\label{sec:PDE_c}
For the second type of non-trivial interaction, a chosen movement event takes place from the current site if, and only if, the target site and all the intermediate sites are available (see Fig. \hspace{-0.3em }\ref{fig:schematic_exclusion}). This leads to the following occupancy master equation for the right-moving subpopulation: 
\begin{align}
\label{eq:ME-2}
\begin{split}	
R^{t+\tau}_{i,j}&=R^t_{i,j}\\&+\frac{\tau P_m}{4}\Big[ (1+\varphi )R^t_{i-v,j} \prod_{s=0}^{v-1} \left(1-C^t_{i-s,j}\right) + (1-\varphi)R^t_{i+v,j} \prod_{s=0}^{v-1} \left(1-C^t_{i+s,j}\right)\\ 
	&+ R^t_{i,j+v}\prod_{s=0}^{v-1} \left(1-C^t_{i,j+s}\right)+R^t_{i,j-v}\prod_{s=0}^{v-1} \left(1-C^t_{i,j-s}\right) \Big] 
	 -\frac{\tau P_m}{4} R^t_{i,j} \Big[ (1+\varphi ) \prod_{s=1}^v \left(1-C^t_{i+s,j}\right)\\&+ (1-\varphi)\prod_{s=1}^v \left(1-C^t_{i-s,j}\right) +  \prod_{s=1}^v \left(1-C^t_{i,j+s}\right) +\prod_{s=1}^v \left(1-C^t_{i,j-s}\right)\Big]\\
	&+ \frac{\tau P_r}{4} \left[ L^t_{i,j}+U^t_{i,j}+D^t_{i,j}-3R^t_{i,j}\right] +\mathcal{O}(\tau^2).
	\end{split}
	\end{align} 
 Again, we refer the reader to the Appendix (see system \hspace{-0.2em }\eqref{eq:master_2} in Appendix \ref{sec:OME}) for the other three occupancy master equations. Upon Taylor expansion and taking the appropriate limits, as before, we obtain:
\begin{equation}
\begin{split}
	\label{eq:PDE_2}
	%equation for R
	&\begin{aligned}
	\D{R}{t}=&v^2\mu \left[R \nabla \left( (1-C)^{v-1} \nabla C\right)+(1-C)\nabla \left( (1-C)^{v-1} \nabla R\right)\right]\\& - v\nu \D{}{x} \left[ R(1-C)^v\right]+\frac{P_r}{4}(C -4 R) 
	,\end{aligned}
	\end{split}
\end{equation}
where $\mu$ and $\nu$ are defined in equations \eqref{eq:coeff}. See system \eqref{eq:PDE_2} of the Appendix \ref{sec:CDS} for the full set of equations. The main difference between equations \eqref{eq:PDE_2} in comparison to the previous exclusion type, characterised by equation \eqref{eq:PDE_1}, is that the rescaling factor, which accounts for the crowding effect, now depends on the $v$-th power of the total density. Notice that for $v>1$,  by adding the equations for the four subpopulations, we obtain non-linear diffusion for the total cell density:
\begin{equation}
	\begin{aligned}
	\label{eq:tot_PDE2}
	\D{C}{t}=&v^2\mu \nabla \left((1-C)^{v-1}\nabla C\right) \\
	&+ v\nu \D{}{y} \left[ (D-U)(1-C)^v\right]+ v\nu \D{}{x} \left[ (L-R)(1-C)^v\right]  \; .
	\end{aligned}
	\end{equation}	
This suggests that the  increase in aborted movements at the micro-scale in these type 2 interactions affects the long-term diffusive behaviour.  

\subsection*{Type 3 interaction}
\label{sec:PDE_d}
Finally, we consider the third (non-trivial) and most mathematically complex form of interaction. This consists of a focal agent moving to the furthest available site in its path towards the target site before (potentially) being blocked. The occupancy master equation for the right-subpopulation reads:
\begin{align}
\label{eq:ME-3}	
\begin{split}
R^{t+\tau}_{i,j}&=R^t_{i,j}\\&+\frac{\tau P_m}{4}\Big[ (1+\varphi )R^t_{i-v,j} \prod_{s=0}^{v-1} \left(1-C^t_{i-s,j}\right) + (1-\varphi)R^t_{i+v,j} \prod_{s=0}^{v-1} \left(1-C^t_{i+s,j}\right)\\ 
	&+ R^t_{i,j+v}\prod_{s=0}^{v-1} \left(1-C^t_{i,j+s}\right)+R^t_{i,j-v}\prod_{s=0}^{v-1} \left(1-C^t_{i,j-s}\right) \Big] 
	 -\frac{\tau P_m}{4} R^t_{i,j} \Big[ (1+\varphi ) \prod_{s=1}^v \left(1-C^t_{i+s,j}\right)\\&+ (1-\varphi)\prod_{s=1}^v \left(1-C^t_{i-s,j}\right) +  \prod_{s=1}^v \left(1-C^t_{i,j+s}\right) +\prod_{s=1}^v \left(1-C^t_{i,j-s}\right)\Big]\\
	& +\frac{\tau P_m}{4}\Big[ (1+\varphi )C^t_{i+1,j} \sum_{k=1}^{v-1} R^t_{i-k,j} \prod_{s=0}^{k-1} \left(1-C^t_{i-s,j}\right) + (1-\varphi)C^t_{i-1,j} \sum_{k=1}^{v-1} R^t_{i+k,j} \prod_{s=0}^{k-1} \left(1-C^t_{i+s,j}\right)\\ 
	&+ C^t_{i,j-1} \sum_{k=1}^{v-1} R^t_{i,j+k}\prod_{s=0}^{k-1} \left(1-C^t_{i,j+s}\right)+C^t_{i,j+1} \sum_{k=1}^{v-1} R^t_{i,j-k}\prod_{s=0}^{k-1} \left(1-C^t_{i,j-s}\right) \Big] \\ 
	&-\frac{\tau P_m}{4} R^t_{i,j} \Big[ (1+\varphi ) \sum_{k=2}^v \prod_{s=1}^{k-1} C^t_{i+k,j} \left(1-C^t_{i+s,j}\right)+ (1-\varphi)\sum_{k=1}^v \prod_{s=1}^{k-1} C^t_{i-k,j} \left(1-C^t_{i-s,j}\right) \\ 
	&+ \sum_{k=2}^v \prod_{s=1}^{k-1} C^t_{i,j+k} \left(1-C^t_{i,j+s}\right) +\sum_{k=2}^v\prod_{s=1}^{k-1} C^t_{i,j-k} \left(1-C^t_{i,j-s}\right)\Big]\\
	&+ \frac{\tau P_r}{4} \left[ L^t_{i,j}+U^t_{i,j}+D^t_{i,j}-3R^t_{i,j}\right] +\mathcal{O}(\tau^2).
	\end{split}
\end{align}
See system \eqref{eq:master_3} in Appendix \ref{sec:OME} for the corresponding occupancy master equations for the other subpopulations. By Taylor expanding and taking the appropriate limits, as before, we obtain the continuum approximation given by:
\begin{equation}
\begin{split}
	\label{eq:PDE_3}
	%equation for R
	&\begin{aligned}
	\D{R}{t}=&\mu \nabla \left[\sum_{k=1}^v (1-C)^{k-1}  \left[ (2k-1)(1-C) \nabla R -k (k-2) R \nabla C \right] \right] \\& + \nu \D{}{x} \left[ \frac{(1-C)\left( (1-C)^v -1\right) }{C} R\right] +\frac{P_r}{4}(C -4 R) 
	,\end{aligned} \\
\end{split}
\end{equation}
where $\mu$ and $\nu$ are as defined in system \eqref{eq:coeff}. The full set of equations is given in Appendix \ref{sec:CDS}. As with type 2 interactions, the polynomial rescaling factor due to the volume exclusion is of order $v$.  Notice that the advective terms contain a factor $C$ in the denominator. We choose to write the advective coefficients this way for notational convenience. Upon expansion of the numerator we see that it also contains a factor $C$, which cancels with the denominator, demonstrating that the coefficient, when simplified, is a polynomial, rather than a quotient. The diffusive terms comprise a sum over $k=1, \dots, v$, which  reflects the possible movement events of length $k$. As for the previous case, by adding all the diffusive terms together for the four subpopulations, we obtain non-linear diffusion for the total population:
\begin{equation}
	\begin{aligned}
	\label{eq:tot_PDE3}
	\D{C}{t}=&\mu \nabla \left[\sum_{k=1}^v (1-C)^{k-1}  \left[ (2k-1) \nabla C +(1-k^2) C \nabla C \right] \right] \\& + \nu \D{}{x} \left[ \frac{(1-C)\left( (1-C)^v -1\right) }{C} (R-L)\right]\\
	& + \nu \D{}{y} \left[ \frac{(1-C)\left( (1-C)^v -1\right) }{C} (U-D)\right].
	\end{aligned}
	\end{equation}	
As expected, the three systems \eqref{eq:PDE_1}, \eqref{eq:PDE_2} and \eqref{eq:PDE_3} for the interacting agent models (type 1, 2 and 3) are equivalent for $v=1$. This is consistent with the ABMs, since the three forms of interactions differ only when movements across multiple lattice sites  are attempted, \textit{i.e.} $v>1$. 

We should mention that the Taylor expansion in equations \eqref{eq:taylor} could be terminated at first order and by following the same steps and taking the limit $\Delta\rightarrow 0$, $\tau\rightarrow 0$ as such that $\Delta/\tau$ is constant, we would have obtained a family of equations similar to systems \eqref{eq:PDE_0}, \eqref{eq:PDE_1}, \eqref{eq:PDE_2} and \eqref{eq:PDE_3} without the contribution of the diffusive terms \cite{treloar2011vjm,treloar2012vjp}. In this case the assumption on the parameter $\varphi$ is no longer necessary. \citet{othmer2000dlt,othmer2002dlt2} studied this type of system in detail in the case of non-interacting agents and for the particular case $\varphi=1$ in one dimension. 
The model of \citet{treloar2011vjm} represents a particular case of the one-dimensional version of the model defined in this paper with $\varphi=1$. \citet{treloar2011vjm} defined their model in terms of the \textit{probabilities} of a single cell of moving and reorienting in a given time step of length $\tau$, which they denote $P$ and $\lambda$, respectively. The reorienting rate, $P_r$, of our models corresponds to the limit $\Lambda= \lim_{\lambda, \tau \rightarrow 0} \lambda/\tau$ of the models of \citet{treloar2011vjm}. Notice, in contrast to the suggestions of \citet{treloar2011vjm}, in our model there are no limitations on the rate $P_r$; it can be chosen to be arbitrary large. \citeauthor{treloar2011vjm} considered a first order Taylor expansion that leads to a system of advective PDEs consistent with our continuum models. Apart from the special case $\lambda=1/2$, for which a simple diffusion equation can be recovered, the nature of their model does not, in general, permit a diffusive limit to be taken. The introduction of the new parameter $\varphi$ in our models allows us to consider a higher order Taylor expansion which results in the diffusive terms in the equations \eqref{eq:PDE_0}, \eqref{eq:PDE_1}, \eqref{eq:PDE_2} and \eqref{eq:PDE_3}.

\section{Results}
\label{sec:results}
In this section we compare the discrete simulations of the ABM with the continuous approximation. Then we investigate how the model behaves under particular choices of the parameters. We reveal three previously unobserved aspects of the model that appear when a high level of persistence is enforced: spike formation, anisotropy and aggregation. Although such phenomena are interesting from a mathematical perspective, they represent potential obstacles for the application of such models to experimental data. We will discuss the implications of such issues and future challenges in Section \ref{sec:conclusion}.

All the ABMs and the corresponding PDEs are simulated on a two-dimensional domain. For the purpose of visualisation, in most examples, we show the results for column-averaged cell density profiles (\textit{i.e.} averaged over the $y$ coordinates).  For these examples, we define the total column-averaged density as:
\begin{equation}
	\label{eq:def_C_bar}
	\bar{C}(x,t)=\frac{1}{L_y}\int_0^{L_y} C(x,y,t) \ud y \, ,
\end{equation}
and $\bar{R}$, $\bar{L}$, $\bar{U}$, $\bar{D}$ correspondingly. In these simulations we choose translationally invariant initial conditions in the vertical direction. In other words, $C(x,y,0)=C(x,0)$ for every $(x,y)\in [1, \Delta L_x] \times [1, \Delta L_y]$. Similarly, the polarised species, $R$, $L$, $U$ and $D$, are also initialised according to a translationally invariant condition. The periodic boundary conditions on the horizontal boundaries imply that translational invariance in the vertical direction is conserved as time evolves, namely:
\begin{equation}
	C(x,y,t)=C(x,t) \; ,
\end{equation}
for every $t\in\mathbb{R}^+$ and for every $(x,y)\in [1, \Delta L_x] \times [1, \Delta L_y]$ and similarly for the four subpopulations $R$, $L$, $U$ and $D$.  

With this in mind, we can now derive the one-dimensional PDEs for the column-averaged densities from the corresponding two-dimensional equations. Formally, this is equivalent to dropping the dependence on $y$ in all the density functions. As an example, we write down the averaged PDEs for the model without interaction (type 0 interactions). We omit the expressions for the other three cases which can be obtained in a similar way. By column averaging the system \eqref{eq:PDE_0}, we obtain the following system of equations

\begin{equation}
\begin{split}
	\label{eq:PDE_0_averaged}
	%equation for R
	&\begin{aligned}
	\D{\bar{R}}{t}=v^2\mu \DD{\bar{R}}{x} - v\nu \D{\bar{R}}{x}+\frac{P_r}{4}(\bar{C} -4 \bar{R}) 
	,\end{aligned} \\
	%equation for L
	&\begin{aligned}
	\D{\bar{L}}{t}=v^2\mu \DD{\bar{L}}{x}+ v\nu \D{\bar{L}}{x}+\frac{P_r}{4}(\bar{C} -4 \bar{L}) 
	,\end{aligned} \\
	%equation for U
	&\begin{aligned}
	\D{\bar{U}}{t}=v^2\mu \DD{\bar{U}}{x}+\frac{P_r}{4}(\bar{C} -4 \bar{U}) 
	,\end{aligned} \\
	%equation for D
	&\begin{aligned}
	\D{\bar{D}}{t}=v^2\mu\DD{\bar{D}}{x}+\frac{P_r}{4}(\bar{C} -4 \bar{D}) 
	,\end{aligned} \\
	\end{split}
\end{equation}
where $\mu$ and $\nu$ are defined as in equations \eqref{eq:coeff}. The boundary conditions, for every $t\in \mathbb{R}^+$, are given by
\begin{equation}
\label{eq:BC_PDE}	
\begin{split}
\D{\bar{R}}{x}(0,t)=\D{\bar{R}}{x}(\Delta L_x,t)=0 \, ,& \qquad 
\D{\bar{L}}{x}(0,t)=\D{\bar{L}}{x}(\Delta L_x,t)=0 \,,\\
\D{\bar{U}}{x}(0,t)=\D{\bar{U}}{x}(\Delta L_x,t)=0 \,,& \qquad
\D{\bar{D}}{x}(0,t)=\D{\bar{D}}{x}(\Delta L_x,t)=0 \, .
\end{split}
\end{equation}\\

\begin{figure}[h!!]
\begin{center}
\subfigure[][]{\includegraphics[width=0.45 \columnwidth]{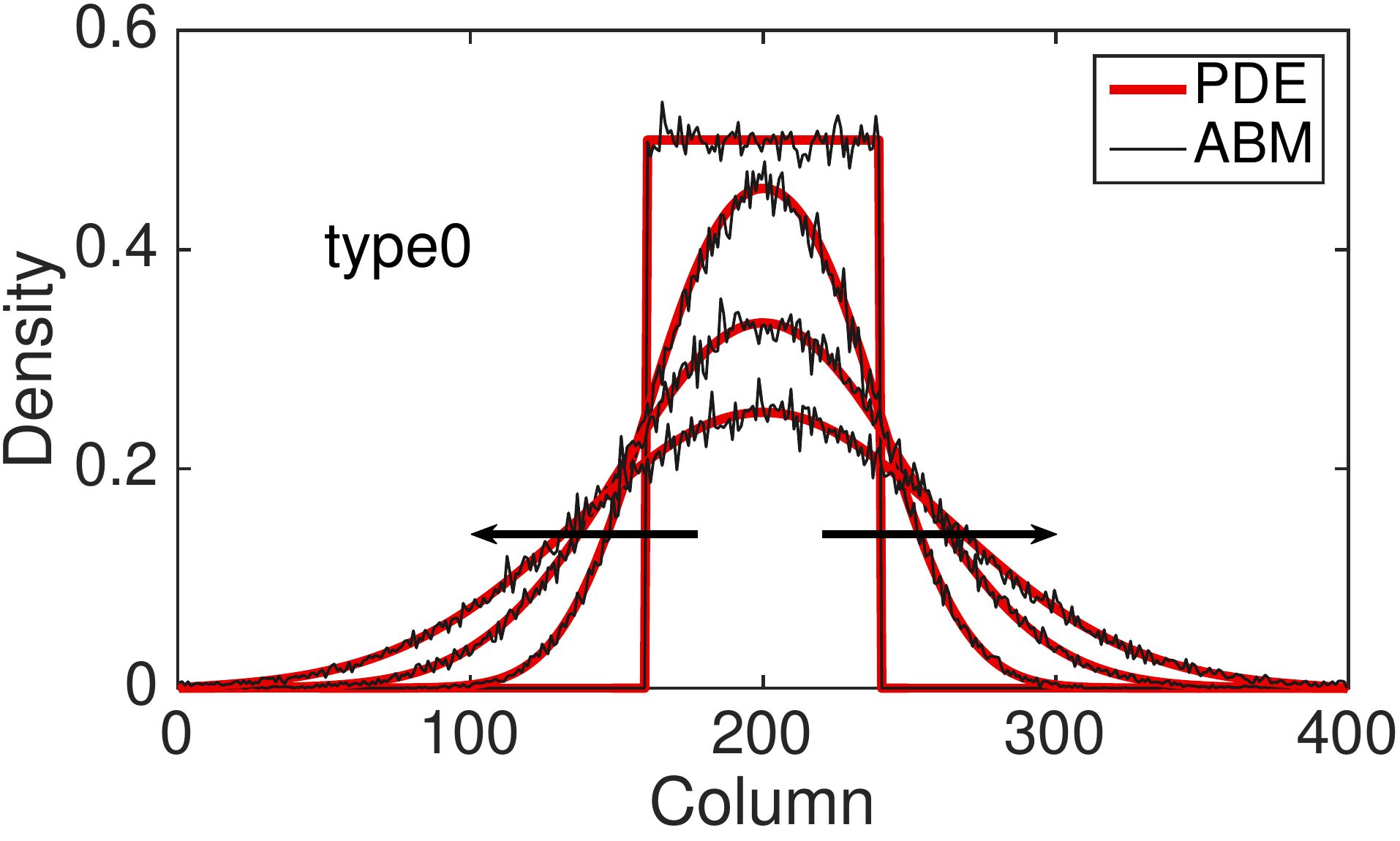}\label{fig:PDE_IBM_type_0} }
\subfigure[][]{\includegraphics[width=0.45 \columnwidth]{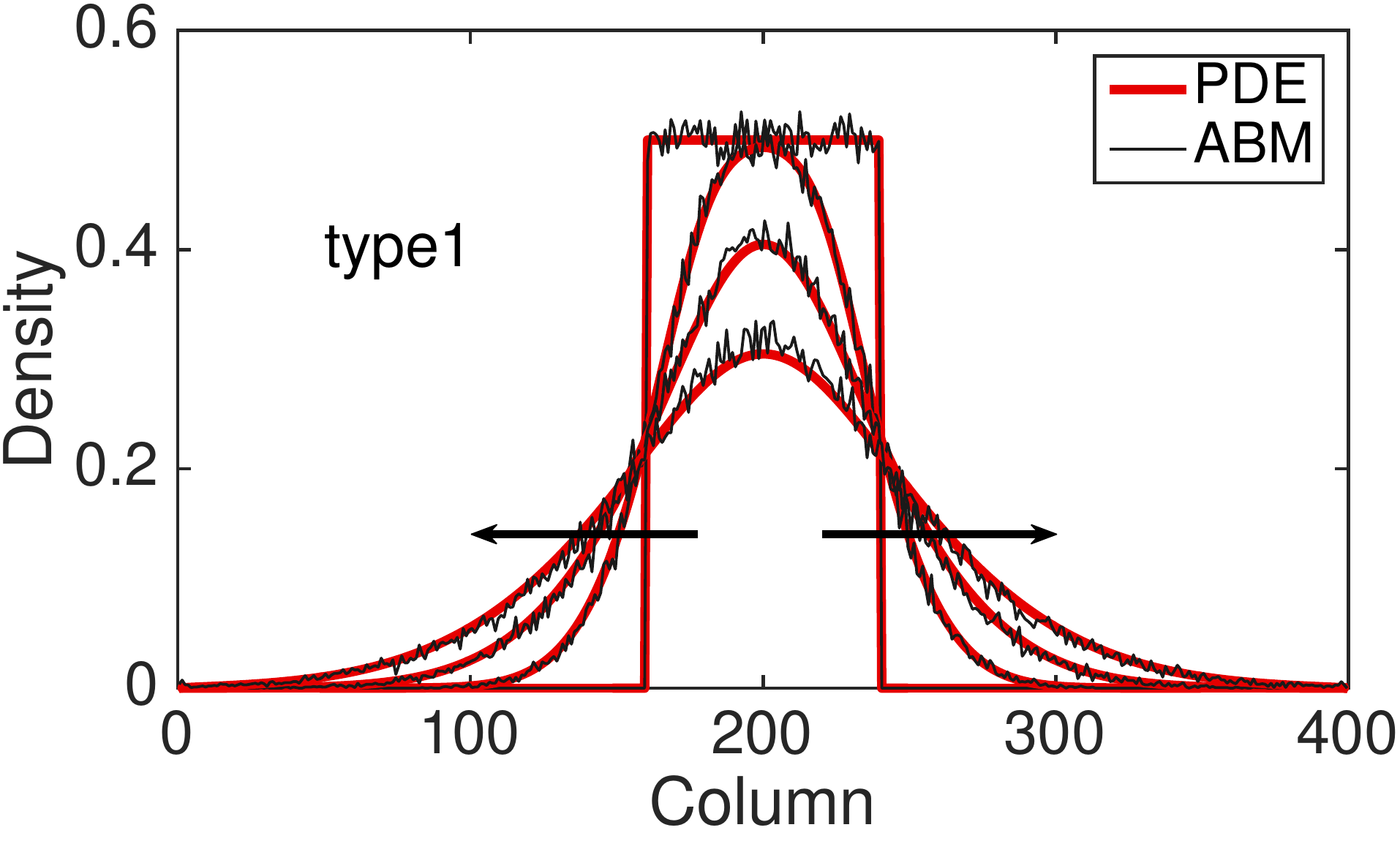}\label{fig:PDE_IBM_type_1} }\\ 
\subfigure[][]{\includegraphics[width=0.45 \columnwidth]{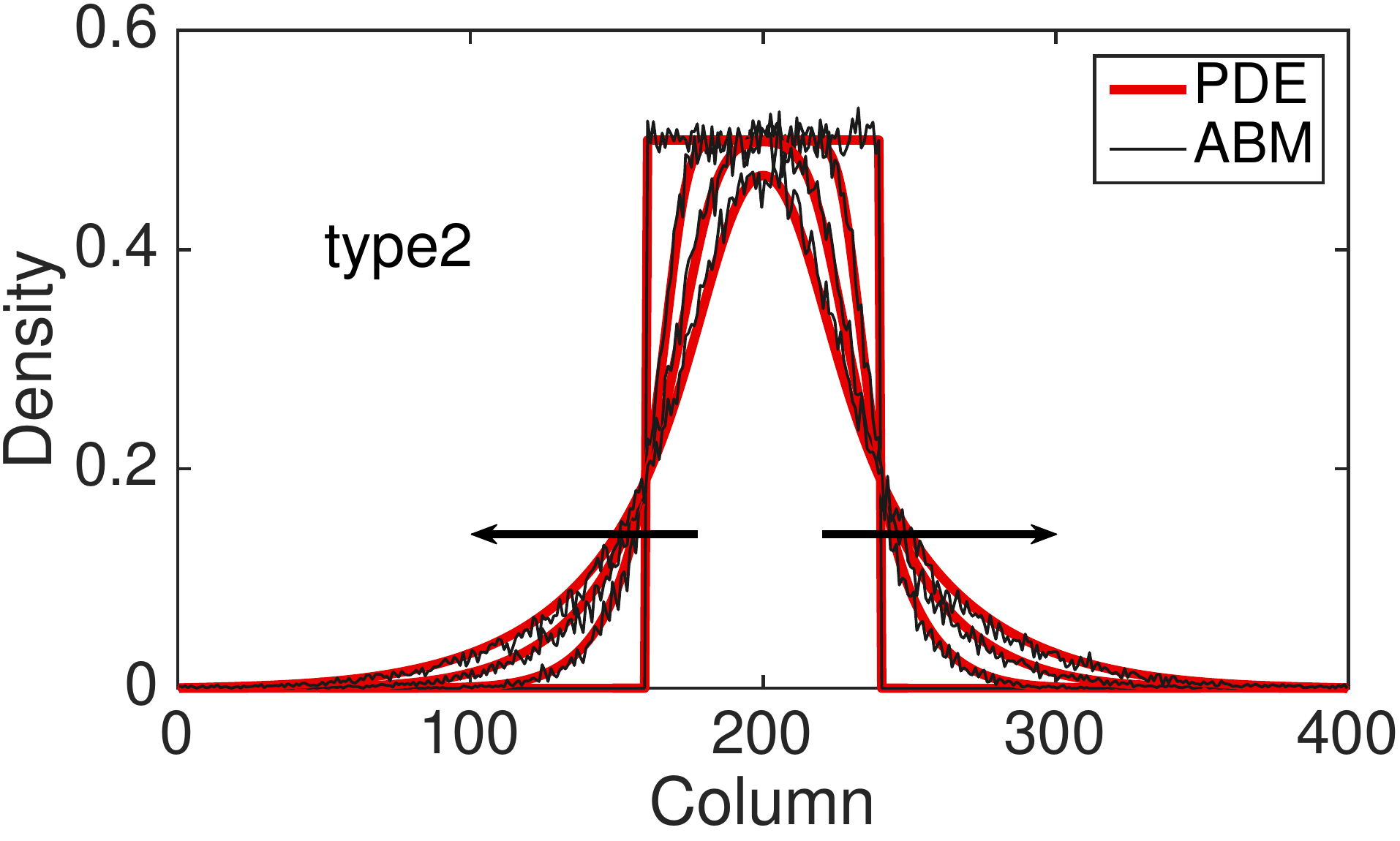}\label{fig:PDE_IBM_type_2} }
\subfigure[][]{\includegraphics[width=0.45 \columnwidth]{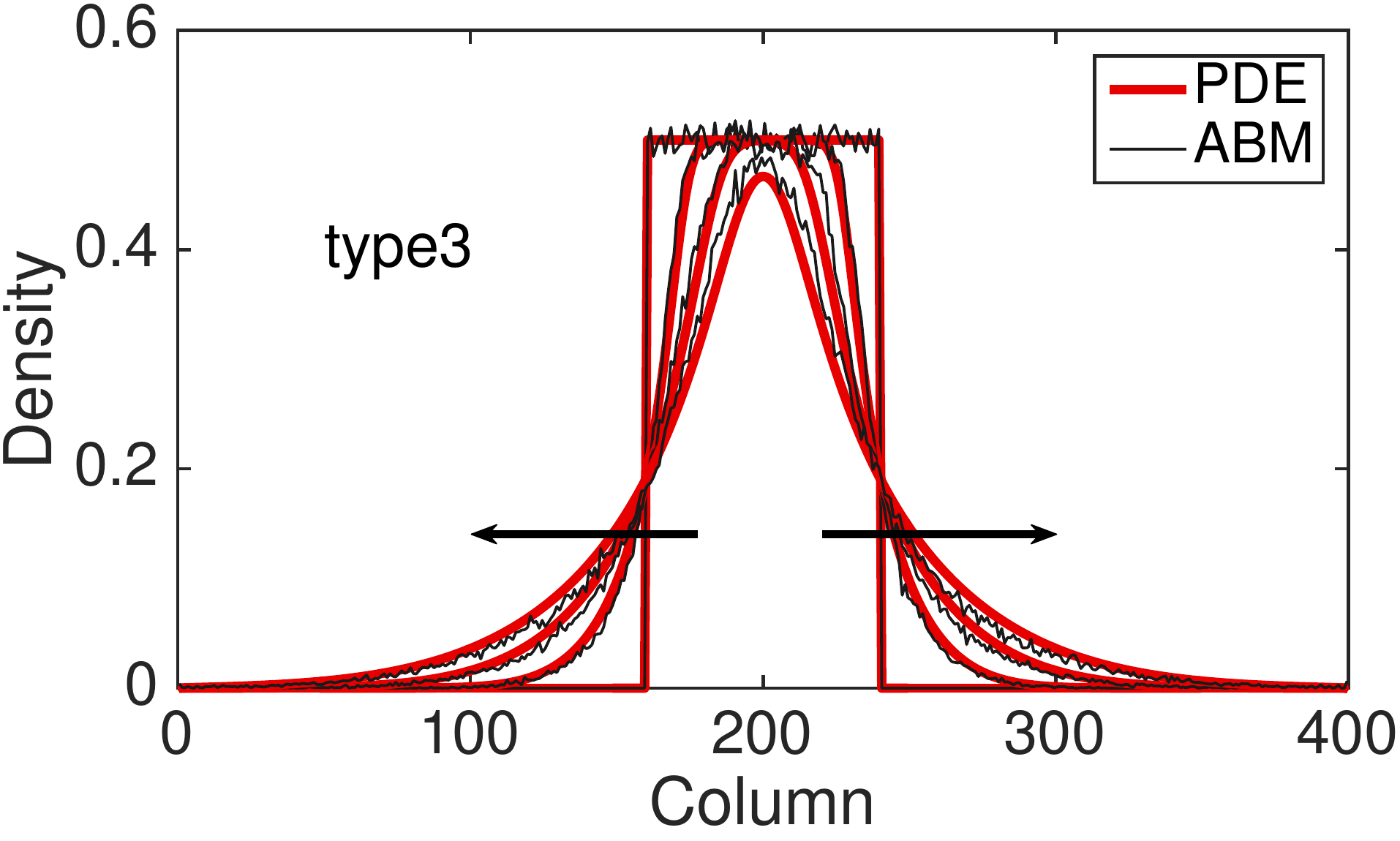}\label{fig:PDE_IBM_type_3} }
\end{center}
\caption{Comparison between ABMs (thin black) and PDEs (thick red) for total column-averaged densities, $\bar{C}(x,t)$, for various forms of interaction. The solutions are displayed at the times $T=0,50,150,300$, with the direction of the black arrows indicating increasing time.  The profiles in panel (a) are for non-interacting agents  (type 0, equation \eqref{eq:PDE_0}). The profiles in panel (b) are for the first type of non-trivial interaction (type 1, equation \eqref{eq:PDE_1}). The profiles in panel (c) are for the second type of non-trivial interaction (type 2, equation \eqref{eq:PDE_2}). The profiles (d) are for the third type of  non-trivial interaction (type 3, equation \eqref{eq:PDE_3}).  All the ABMs are simulated using the Gillespie algorithm \cite{gillespie1977ess} averaged over $M=10$ repeats.}
\label{fig:comparison1} 
\end{figure}

\begin{figure}[h!!!!!!!!!!]
\begin{center}
\includegraphics[width=0.45 \columnwidth]{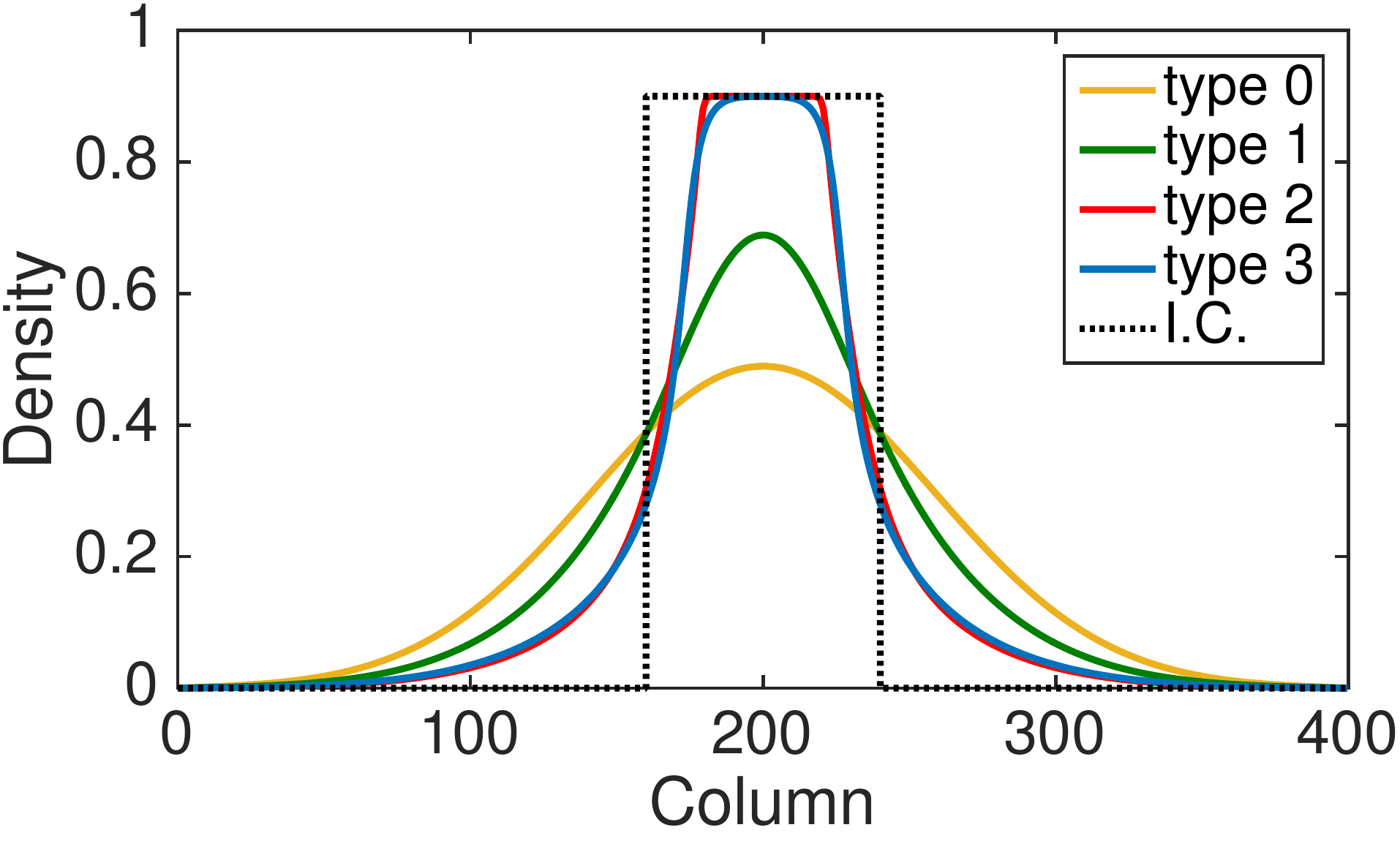}
\end{center}
\caption{Comparison of the numerical solutions of the column-averaged PDEs for the four types of interactions, types 0-3.  The blacked dotted line illustrates the initial condition. For interpretation of the references to colour in this figure legend, the reader is referred to the web version of this article.}
\label{fig:comparison_PDE_250} 
\end{figure}

\begin{figure}[h!!]
\begin{center}
\subfigure[][]{\includegraphics[width=0.45 \columnwidth]{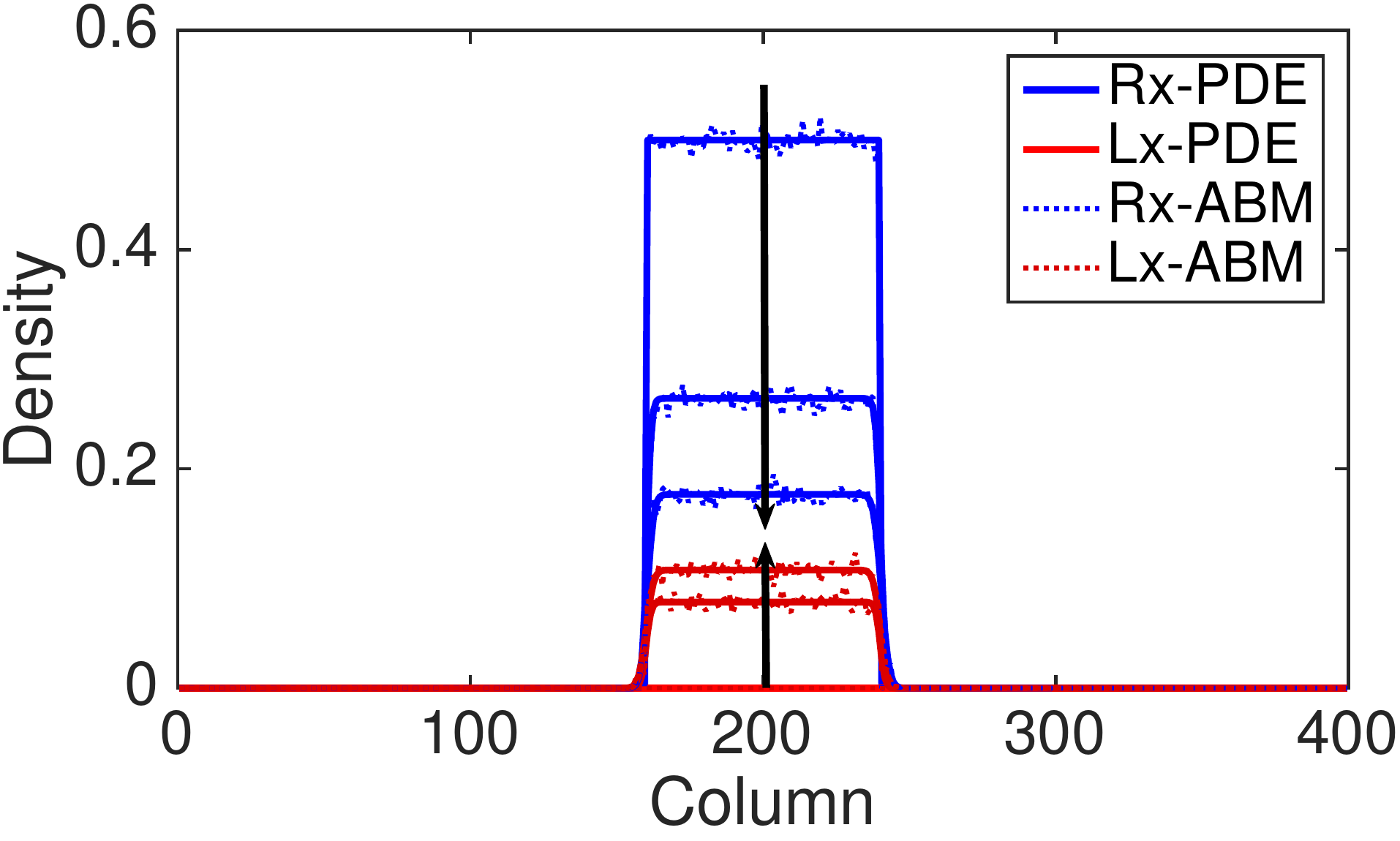} }
\subfigure[][]{\includegraphics[width=0.45 \columnwidth]{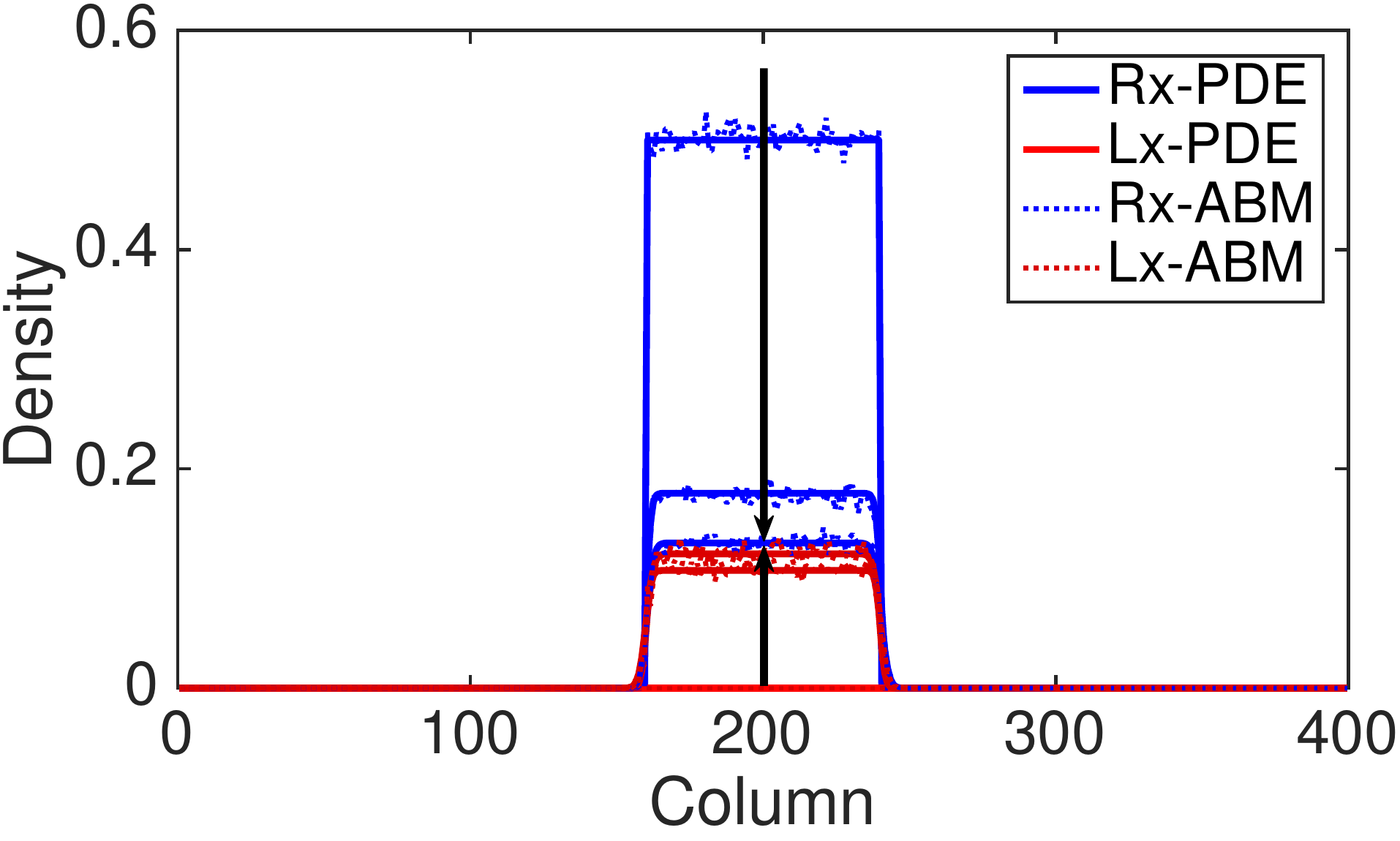} }\\
\end{center}
\caption{Comparison between ABMs (dotted line) and PDEs (full line) of right (blue) and left (red) subpopulations of non-interacting agents (type 0) for large values of reorienting rate $P_r$. In panel (a) $P_r=1$ and in panel (b) $P_r=2$. The simulations are initialised with the central region populated with density $d=0.5$ by only right-polarised cells. The solutions are displayed at the times $T=0,1$ and $2$, with the direction of the black arrows indicating increasing time. The numerical solutions of the PDEs are obtained as described in the text. For interpretation of the references to colour in this figure legend, the reader is referred to the web version of this article.}
\label{fig:comparison_r} 
\end{figure}

\begin{figure}[h!!!!!!!!!!]
\begin{center}
\includegraphics[width=0.45 \columnwidth]{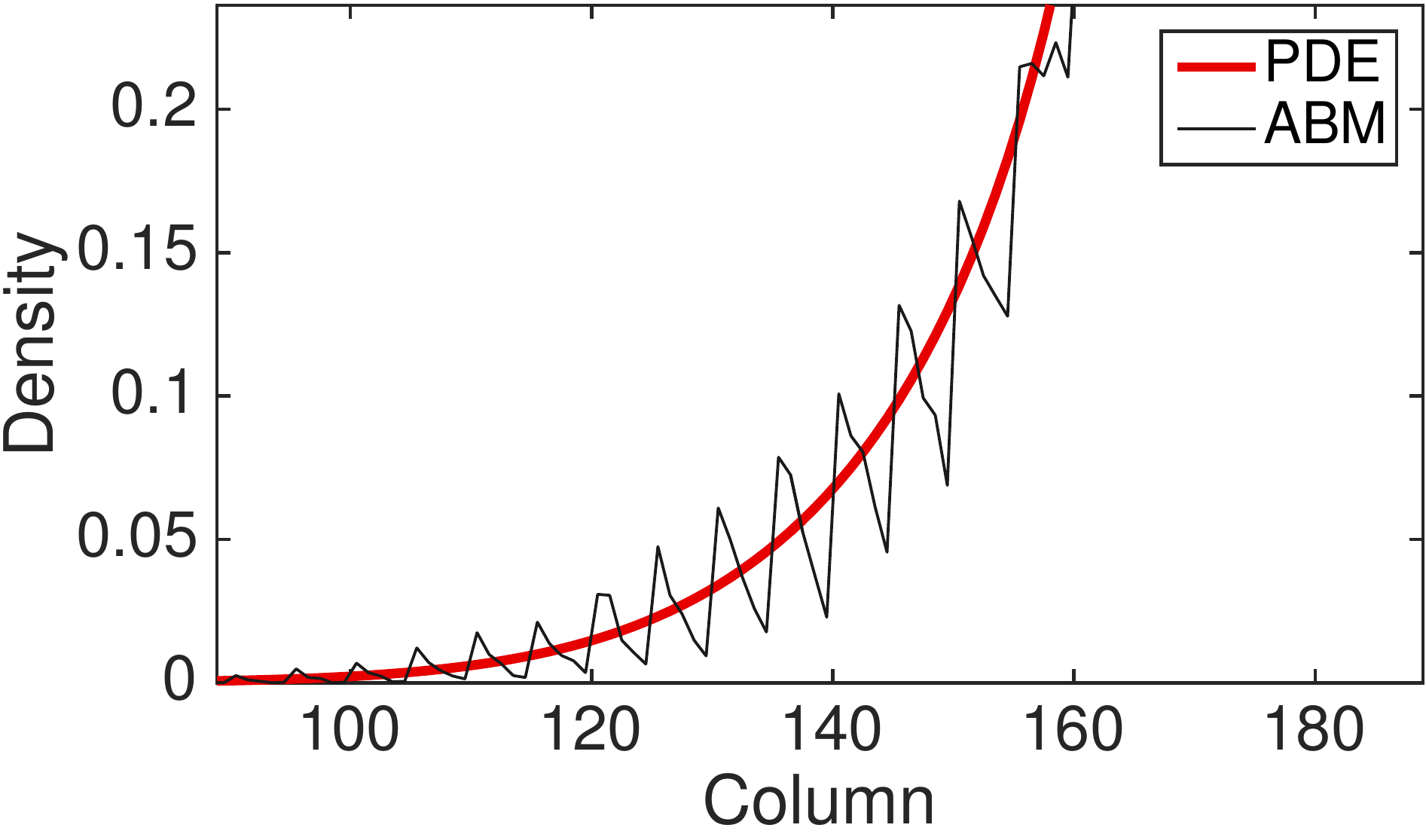}
\end{center}
\caption{A zoomed-in comparison between the column-averaged ABM (thin black) and the PDE (thick red) for $\bar{C}(x,t)$  for the second type of non-trivial interaction (type 2). Density spikes are clearly visible in the ABM, but not in the PDE.}
\label{fig:density_spikes}
\end{figure}

For Figs. \hspace{-0.3em }\ref{fig:comparison1}, \ref{fig:comparison_PDE_250}, \ref{fig:comparison_r} and  \ref{fig:density_spikes} we use the same computational set-up which can be described as follows. The domain is a $400\times 400$ lattice with $\Delta=1$. We impose periodic boundary conditions on the horizontal boundaries and zero-flux boundary conditions on the vertical boundaries. The ABM is simulated using the Gillespie algorithm \cite{gillespie1977ess} for $M=10$ identically prepared repeats. Few repeats are sufficient to compare the mean ABM behaviour to the  solutions of the PDEs for the mean occupancy because column-averaging over 400 rows significantly reduces the noise in the ABM solutions. The numerical solutions of the PDEs are obtained through an implicit Euler method with spatial step $\delta x=0.1$ and time step $\delta t=0.1$, and using Picard iteration with tolerance $\epsilon=10^{-3}$ to solve the non-linear equations.

Fig. \hspace{-0.3em }\ref{fig:comparison1} shows the comparison between the total column-averaged density of the ABM and the PDE, for the four types of interactions at different times.
The system is initialised such that the all sites with $x$ coordinate between $161$ and $240$ are populated uniformly at  at random, with density $d=0.5$. The polarisation of the initial group of cells is chosen uniformly at random, so there is no bias towards any of the four polarisations. The PDE for the total column-averaged density is correspondingly initialised as constant $d=0.5$ in the interval $[160, 240]$, with the density of the individual subpopulations also being constant at $d=0.125$ in this region.  The parameters for the model are $P_m=1$, $P_r=0.2$, $\varphi=0.8$, $v=3$.

\begin{figure}[h!!]
\begin{center}
\subfigure[][]{\includegraphics[width=0.45 \columnwidth]{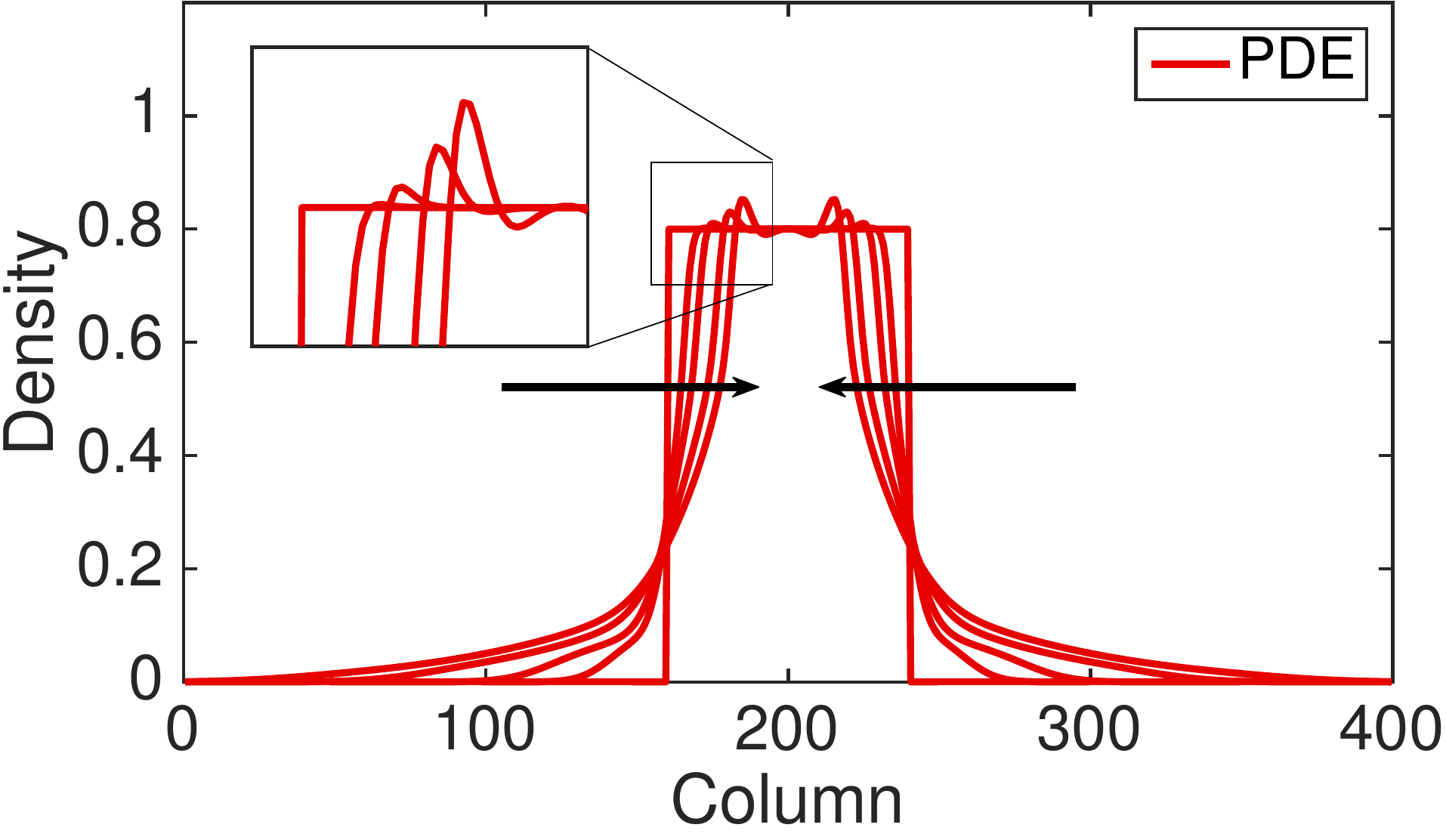} }
\subfigure[][]{\includegraphics[width=0.45 \columnwidth]{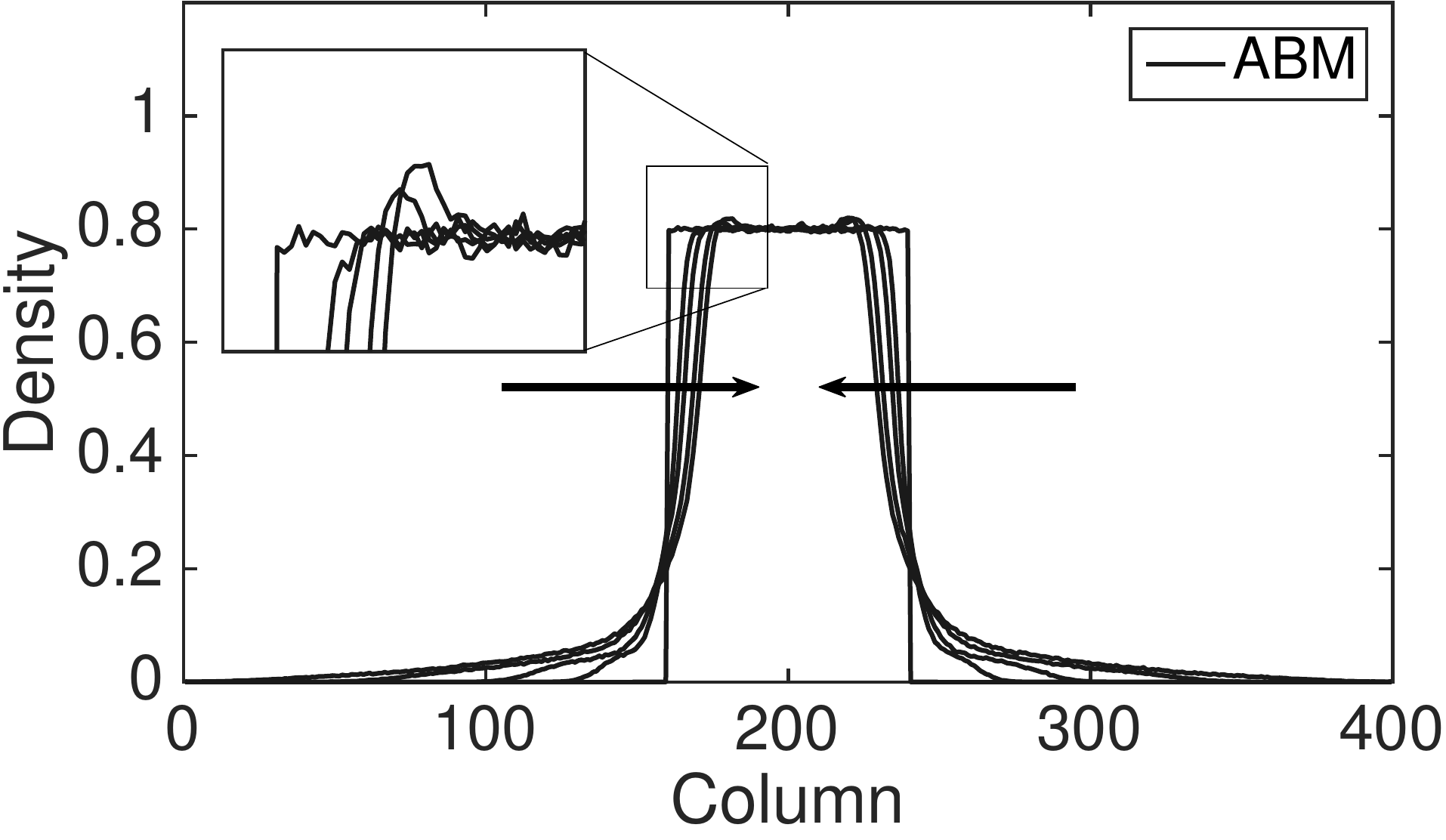} }\\
\subfigure[][]{\includegraphics[width=0.45 \columnwidth]{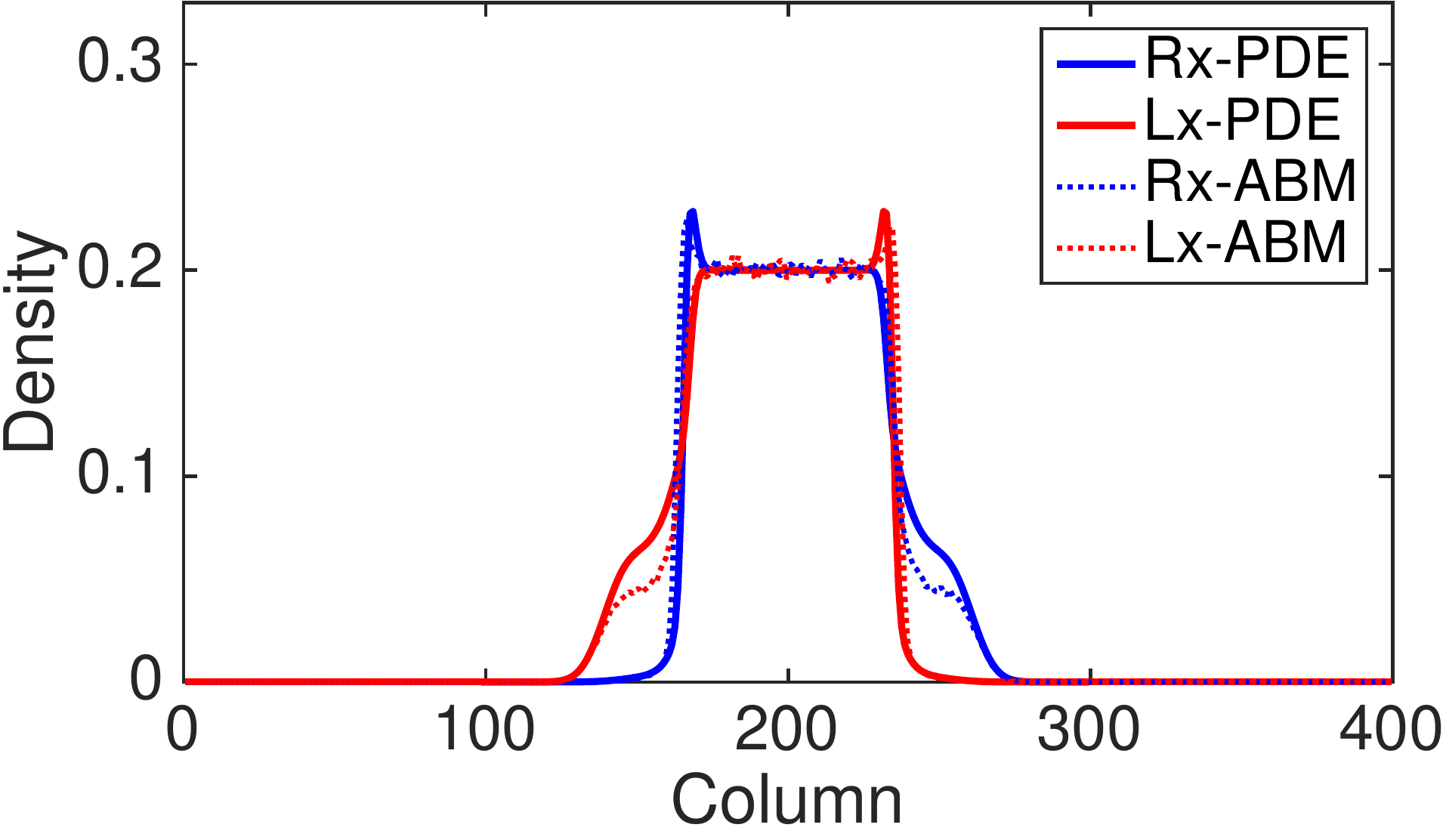}}
\subfigure[][]{\includegraphics[width=0.45 \columnwidth]{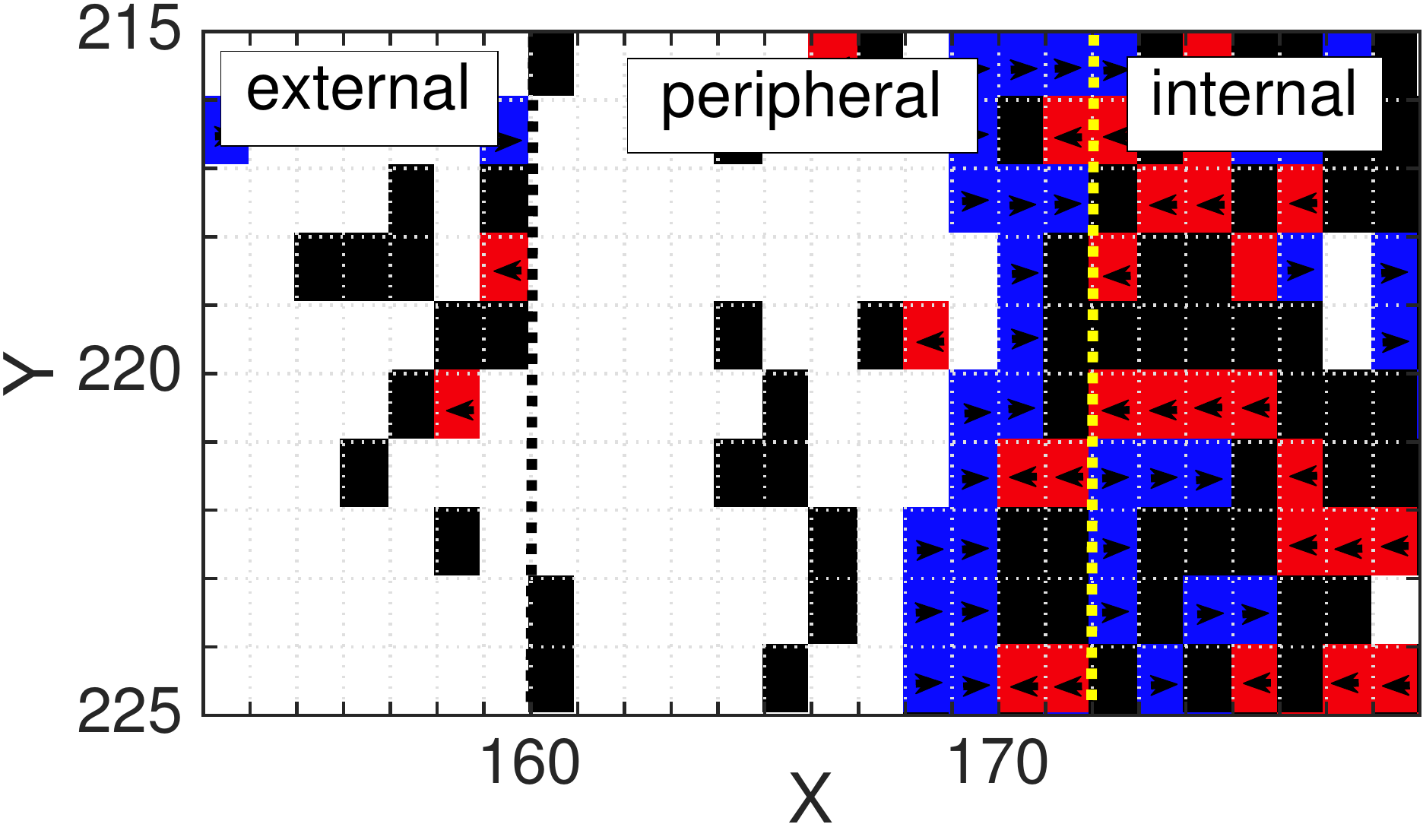}}
\end{center}
\caption{Spontaneous aggregation induced by persistence and volume exclusion. The top panels show (a) the numerical solution of the PDE for the column-averaged total density  and (b) column-averaged density of the ABM, averaged over $M=100$ repeats, for the model with interacting agents (types 1-3 with $v=1$). The profiles are shown at time $T=0,50,100,200,300$ with the direction of the black arrows indicating increasing time. In panel (c) the solution of the PDE (continuous lines) and ABM (dotted lines) are compared for the partial column-averaged densities of right-polarised agents (blue lines) and left-polarised agents (red lines) at time $T=50$. In panel (d) a zoomed-in snapshot of a single simulation of the ABM (types 1-3) is shown at time $T=200$. White sites are empty, blue sites are occupied by a right-polarised agent, red sites are occupied by a left-polarised agent and black sites are occupied by either an up- or down-polarised agent. The black-dotted line represents the border of the initially populated region which divides the external region from the peripheral region, the yellow dotted-line distinguishes between the peripheral and the internal region. For interpretation of the references to colour in this figure legend, the reader is referred to the web version of this article.}
\label{fig:aggregation} 
\end{figure}

 The agreement between the discrete and the continuous descriptions is generally very good for all four types of cell interactions, although we note that discrepancies are most noticeable for the type 3 interactions for which our assumption of independence of site occupancy is least valid. We also tested our continuous approximation for a different parameters values and found that the good agreement with the discrete model holds for a wide parameter range (results not shown). The good agreement is lost, however, as the value of persistence increases, \textit{i.e.} $P_r \ll P_m$ and $\varphi\approx 1$ (see Fig.\,\ref{fig:aggregation}\,(c) for an example). This disagreement is, in part, due to the significant spatial correlations induced by persistence in these models. Agents are highly likely to move to be adjacent to each other rather than aborting their movements (compare type 3 interactions to type 2 interactions, respectively). This tendency is ignored by our continuous approximations. We discuss this issue and potential improvements further in Section \ref{sec:conclusion}.

The hyperbolic nature of previous continuum models of persistence of motion has meant that initial conditions with steep gradients have been difficult to investigate. Due to the diffusive nature of our continuous model, we are now able to examine initial conditions which have steep density gradients and still maintain a good agreement with the discrete model. In particular, this allows us to consider the initial condition described above with only a central region uniformly populated.

To compare the behaviour of the different interaction mechanisms, we increase the initial total density to $d=0.9$ and we display numerical solutions of the four PDEs \eqref{eq:PDE_0}, \eqref{eq:PDE_1}, \eqref{eq:PDE_2} and \eqref{eq:PDE_3} at time $T=250$ (see Fig. \hspace{-0.3em }\ref{fig:comparison_PDE_250}). Non-interacting agents (type 0), lead to a faster agent spreading than any of the interacting types 1-3. Type 1 interactions cause slightly slower spread of agents: although focal agents can jump over their neighbours, a small number of type 1 movement events are aborted if there is a cell in the target site. Type 2 interactions lead to the slowest spreading due to the high proportion of aborted movements events in which the focal agent stays stationary. Agents interacting through the type 3 mechanism spread slightly faster than agents undergoing type 2 interactions: although a significant proportion of movement events are aborted when agents are immediately adjacent to the focal agent, some small movement events are permitted towards a near neighbour which would otherwise be aborted under the implementation of type 2 interactions.

In Fig. \hspace{-0.3em }\ref{fig:comparison_r} we compare the ABM and the continuum approximation for the right- and left-moving subpopulations in order to evidence that the agreement between the models does not only hold at the population-level. The density profiles of the up- and down-moving subpopulations are indistinguishable and we have omitted them for simplicity. However the good of agreement between the discrete and continuum models also holds for these cases (results not shown). We find good agreement, even for large values of the reorienting  parameter, $P_r$, which has previously been thought not to be the case \citep{treloar2011vjm}. Note that the appearance of loss of total mass is due to the fact that we are only visualising two of the four subpopulations in our two-dimensional model.

In the following subsections we outline some of the inherent features of these models which have largely been overlooked, but which must be considered if the model is to be used in real applications.

\paragraph{Density spikes.} 

In Fig. \hspace{-0.3em }\ref{fig:density_spikes} we plot a zoomed-in density profile of the ABM and PDE for the second type of interacting agents (type 2). The parameters and the initial condition are as in Fig. \hspace{-0.3em }\ref{fig:comparison1} apart from $v=5$ and $d=0.8$. The new values of the parameters $v$ and $d$ are chosen in order to highlight the following phenomenon of the ABM. The results reveal a substantial difference between the behaviour of the discrete model, in which regular spikes appear clearly in the density profile, and the behaviour of the continuous model, whose density profile  appears as a smooth function. In other words, the PDE provides correct information on the average number of agents in an interval of length $v$, but it fails to reveal the behaviour of the ABM at smaller scales. Notice that such discrepancy is not due to the stochasticity of the ABM, but it is a systematic feature of the model at the agent-level. In order to gain an intuition of how the spikes appear in the ABM, notice that the agents which first leave the initial region (the region of the domain in which agents are initiated) by making a long jump of the maximum distance, $v=5$, form an effective barrier for the following agents. Subsequent agents leaving the initial region accumulate behind this barrier. This mechanism repeats itself as the furthest agents jump again producing a second effective barrier at distance $\Delta v=5$.  This mechanism produces a density profile characterised by multiple spikes at distances which are multiples of $v$. The amplitude of the spikes decreases with distance from the initial condition as some of the barrier agents or their successors change orientation or become less synchronised in their outward movements leading to successively more porous barriers. The same behaviour is also observed in the other form of complex interaction (type 3, results not shown). This departure from the PDE model becomes more evident as $v$ increases. 

\paragraph{Anisotropy.} 
\label{sec:anisotropy}
\begin{figure}[h!!!!!!!]
\begin{center}
\subfigure[][]{\includegraphics[width=0.4 \columnwidth]{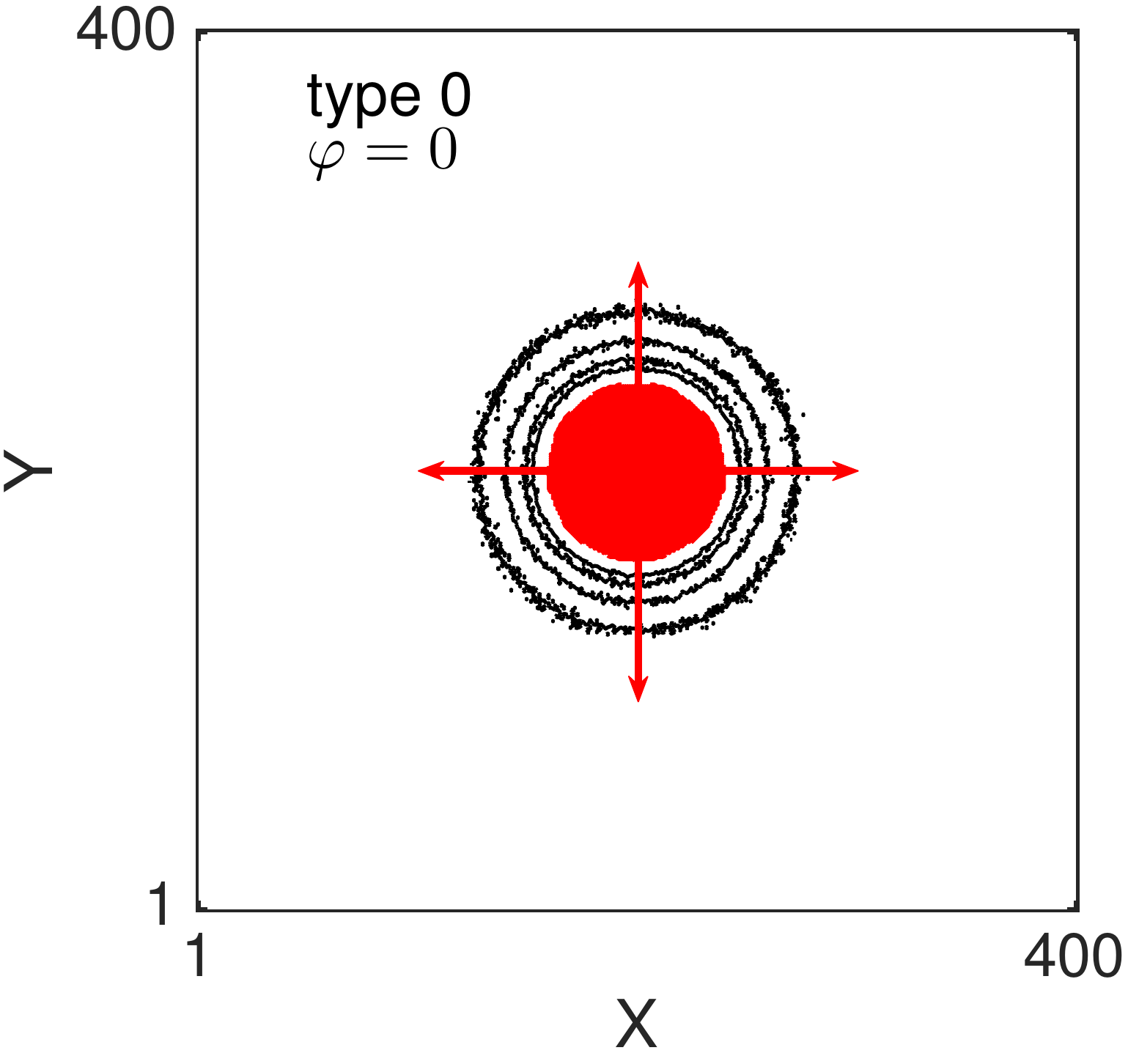} }
\subfigure[][]{\includegraphics[width=0.4 \columnwidth]{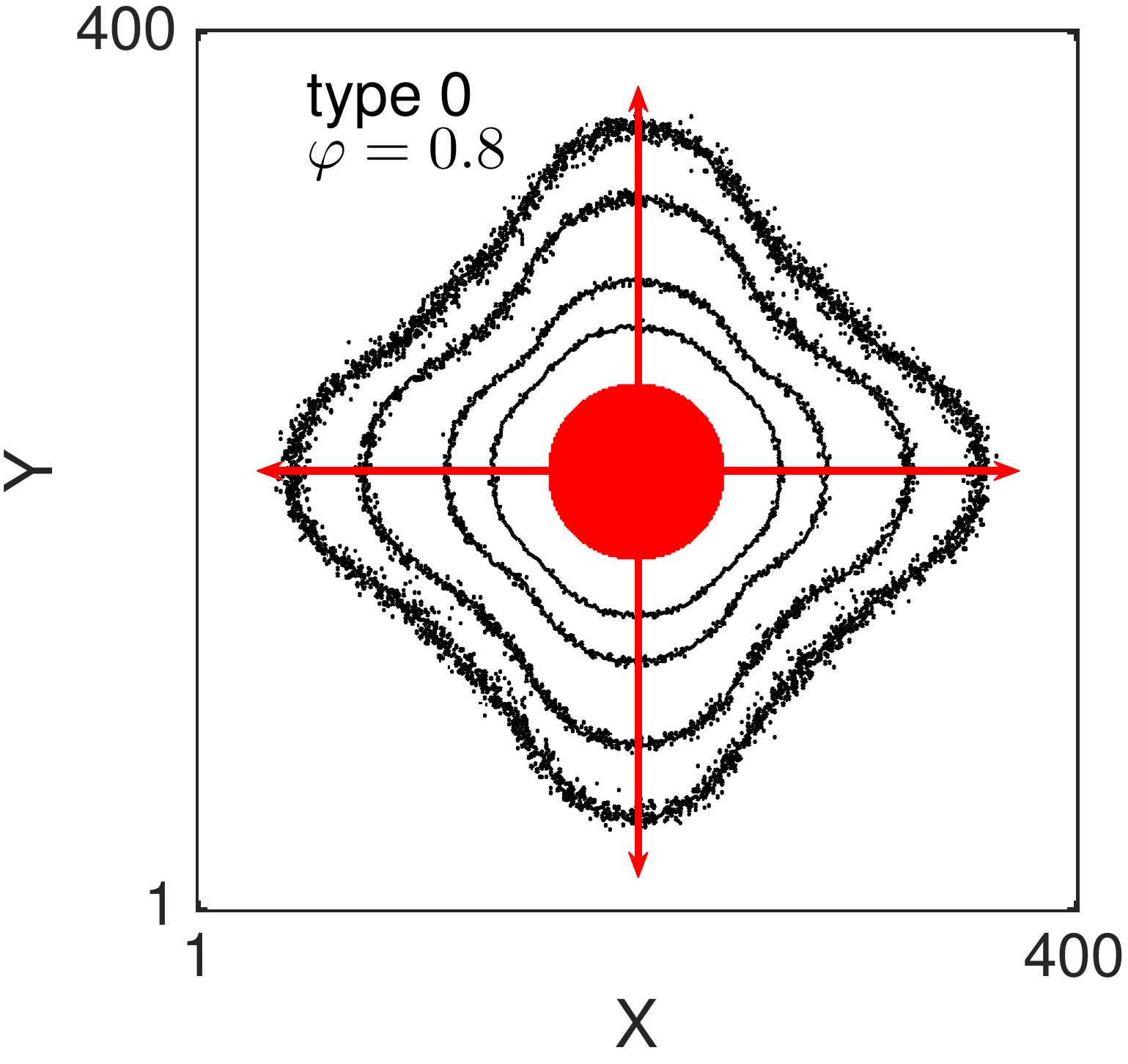} }\\
\subfigure[][]{\includegraphics[width=0.4 \columnwidth]{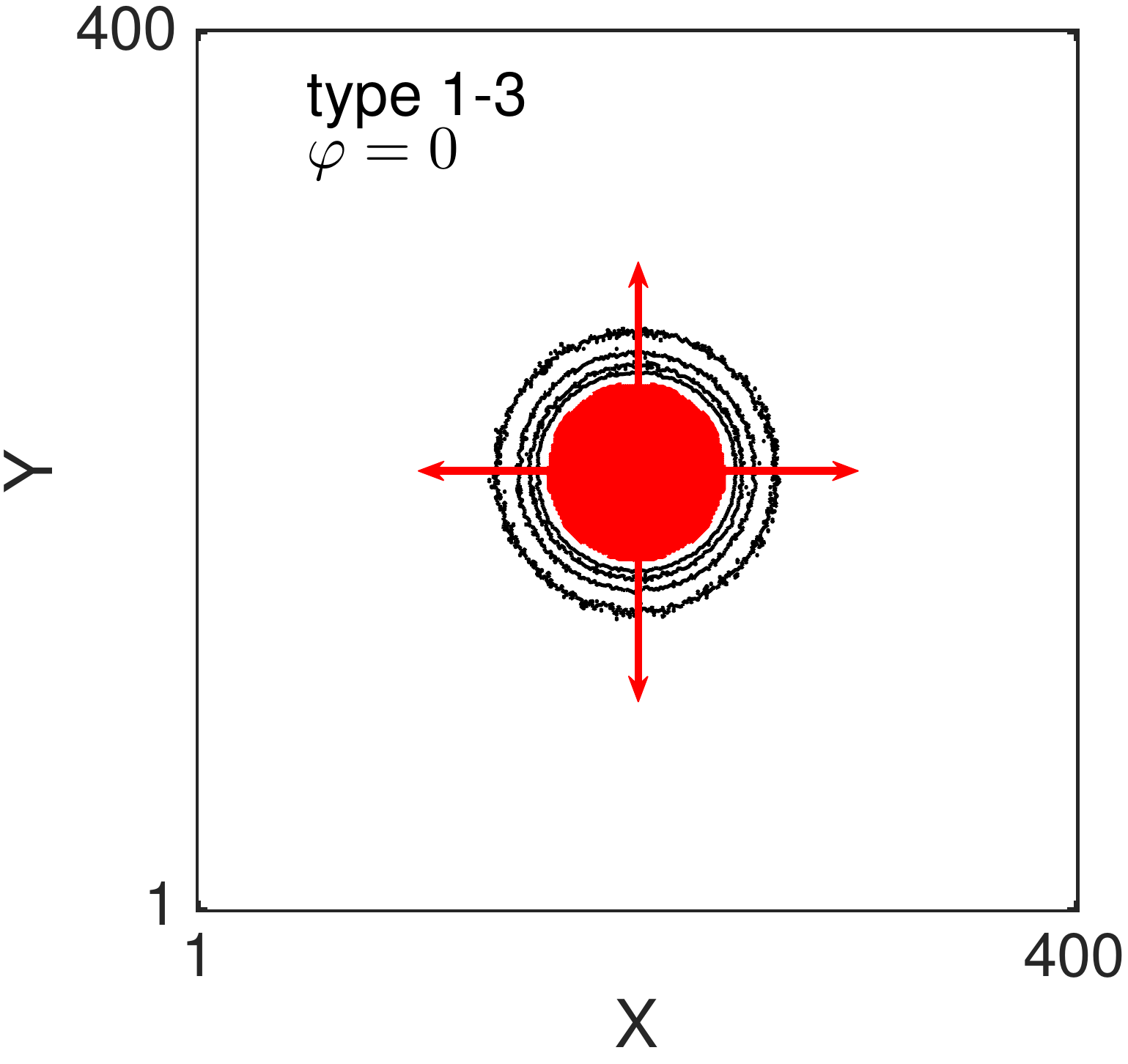} }
\subfigure[][]{\includegraphics[width=0.4 \columnwidth]{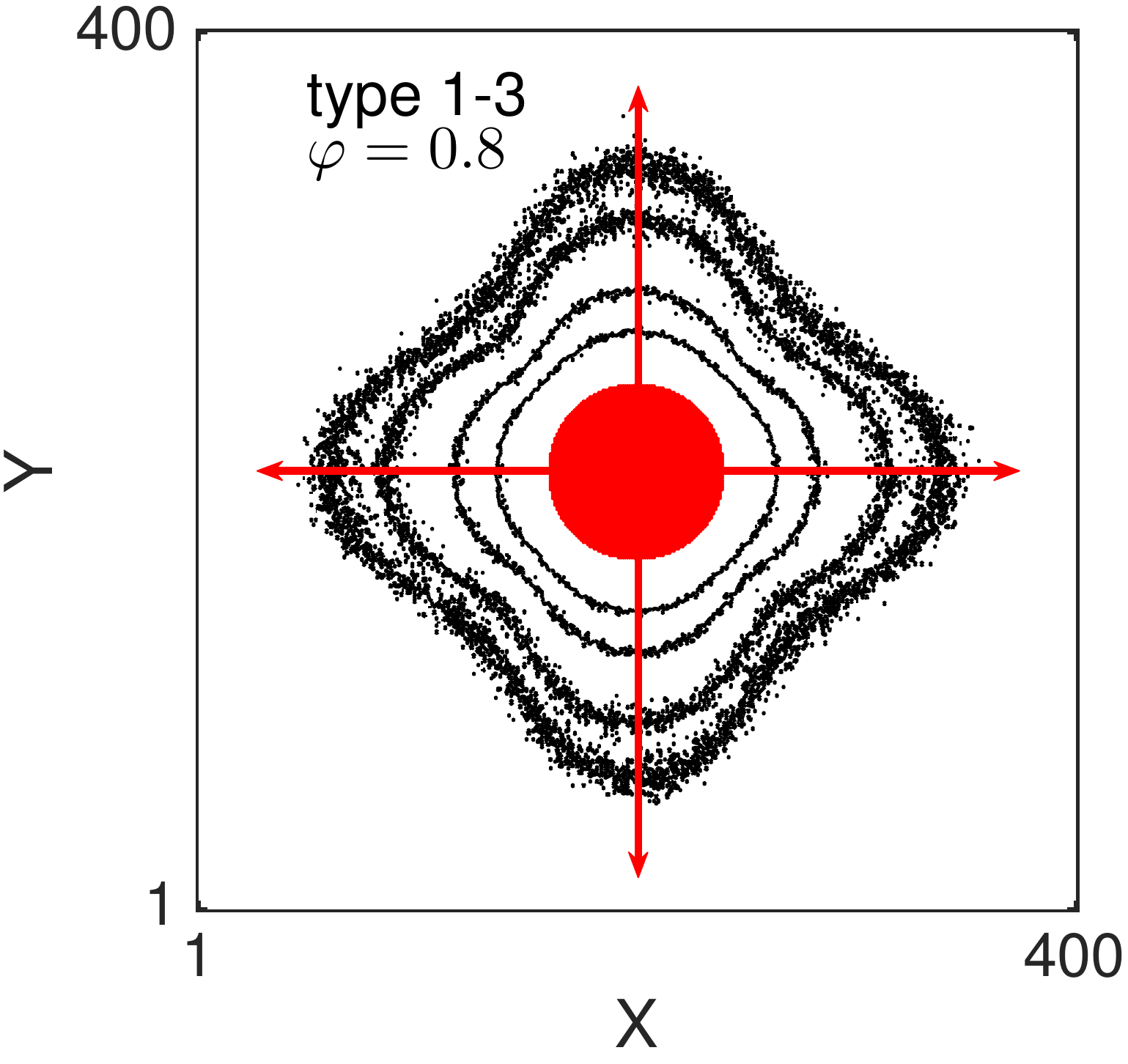} }
\end{center}
\caption{Anisotropy of the ABM in two dimensions. The black contour lines represent values of equal total density, $C^T_{i,j}=0.01$, at times $T=50,100,200,300$ with time increasing in the direction of the red lines. Panels (a) and (b) display non-interacting agent profiles (type  0): panel (a) without persistence of motion, $\varphi=0$, and panel (b) with strong persistence of motion, $\varphi=0.8$. Panels (c) and (d) display agent profiles for the non-trivial agent interactions (types 1-3): panel (a) without persistence, $\varphi=0$, and panel (b) with strong persistence, $\varphi=0.8$.}
\label{fig:anisotropy} 
\end{figure}
One of the key feature of all of our models, for large value of persistence,  is the appearance of (positive) anisotropy in the axial directions \cite{thompson2011lmn}. In Fig. \hspace{-0.3em }\ref{fig:anisotropy} we show the two-dimensional density contour lines $C^T_{i,j}=0.01$ of the ABMs (averaged over $M=1000$ realisations) evolve as time increases. All the ABM simulations are initialised by populating a central circular region of diameter $r=40$ (i.e. the sites whose centres lie less than $40\Delta$ away from the centre of the domain). The contour density lines are recorded at times $T= 50,100,200,300$. In all the examples we choose $P_m=1$, $P_r=0.01$ and $v=1$. Notice that with this choice of parameters the three types of volume-excluding interactions (types 1-3) are equivalent. We repeat the simulations for non-interacting agents (Fig. \hspace{-0.3em }\ref{fig:anisotropy} (a) and (b)) and for interacting agents (Fig. \hspace{-0.3em }\ref{fig:anisotropy} (c) and (d)). For each scenario we consider the non-persistent case, $\varphi=0$, (Fig. \hspace{-0.3em }\ref{fig:anisotropy} (a) and (c)) and the strongly persistent case, $\varphi=0.8$, (Fig. \hspace{-0.3em }\ref{fig:anisotropy} (b) and (d)). When persistence is not included, the dynamics of the agents correspond to simple excluding walks. In this case the isotropy of the initial condition is known to be preserved as the system evolves (Fig. \hspace{-0.3em }\ref{fig:anisotropy} (a) and (c)) \cite{murray2007mbi, deutsch2007cam, codling2008rwm}. In particular, the density profiles conserve the circular shape of the initial region, meaning that there is no preferential direction of migration. When the persistence is switched on (Fig. \hspace{-0.3em }\ref{fig:anisotropy} (b) and (d)) the isotropy is lost. Cells spread faster in the four axial directions (red arrows), due to their polarisations, and this leads to density contour lines with a ``diamond'' shape. 

It should be noticed that this phenomenon is not produced by the mechanism of jumping multiple lattice steps simultaneously, in fact we deliberately chose $v=1$ to illustrate this. The anisotropy is, instead, an intrinsic feature of the persistent model combined with the lattice environment. Such anisotropic behaviour has not been observed in previous studies, which focused on the one-dimensional scenario \cite{treloar2011vjm,treloar2012vjp}, because the higher-dimensional setup is a necessary condition for the anisotropy to appear. 
othmer1988mdb
\paragraph{Spontaneous aggregation.}
\label{sec:aggregation}
The last phenomenon that we highlight here is the emergence of short-term aggregation in density profiles driven by the interplay of persistence and volume exclusion. In Fig. \hspace{-0.3em }\ref{fig:aggregation} (a) and (b) we show the total density of the column-averaged PDE and ABM, respectively, for the model with agent interaction (types 1-3) and parameters $P_m=1$, $P_r=0.01$, $\varphi=0.9$, $v=1$. Such a choice of parameters leads to strong persistence. This is needed in order to make the aggregation phenomenon evident. Additionally, we increased the number of repeats, to $M=100$, to reduce the noise and better demonstrate the phenomenon in the ABM averaged profiles. The profiles are shown at times $T=0,50,100,200,300$. The plots show a snapshot of the system in which total density towards the edge of the initially populated region increases above the initial value producing a spontaneous non-monotonicity (travelling outwards from the centre of the domain in either direction) in the density profile towards the edge of the initial interval. A closer look at the density for right and left polarised agents at time $T=50$ (see Fig. \hspace{-0.3em }\ref{fig:aggregation} (c)) reveals that the increment on the left-hand side of the initial condition is caused by an accumulation of right-polarised cells and \textit{vice versa} for the other side. Although in the early stages ($T=50$) the accumulation is visible only in the profiles of the differently polarised populations (see Fig. \hspace{-0.3em }\ref{fig:aggregation} (c)), eventually ($T=200$) the non-monotonicity appears at the total population-level (see Figs. \hspace{-0.3em }\ref{fig:aggregation} (a) and (b)). 

In order to explain this phenomenon of \textit{spontaneous aggregation} at the microscopic level, we consider a scenario with initially high density and high values of persistence ($P_r \ll 1 $ and $\varphi \approx 1$). In Fig. \hspace{-0.3em }\ref{fig:aggregation} (d) we display at portion of a single ABM simulation  magnified in the region around the formation of the left peak. We partition the figure in three regions: \textit{external}, which corresponds to the region outside the initially populated region and that is empty at time $t=0$; \textit{peripheral}, which represents the region containing the border of the initially populated region and where the aggregation takes place; and \textit{internal}, which represents the centre of the initially populated region. In the first phase of the simulation, the right-polarised cells in the peripheral region are blocked by the high density in the internal region and they are likely to remain in their position unless they reorient, which happens with low probability. Meanwhile, some of the other polarised cells which occupy the peripheral region, spread into the external region on the left-hand side which creates a decrease in the total density of the peripheral region. This allows the right-polarised cells in the peripheral region to move rightwards, further into the peripheral region, to aggregate and hence to form a barrier for the cells in the internal region (see blue squares in the proximity of the yellow dotted line in Fig. \hspace{-0.3em }\ref{fig:aggregation} (d), for example). The agents in the internal region remain trapped by this obstruction. Notably, we can also see a weak form of aggregation on the internal sides of the two barriers in the PDE density profiles in Fig. \hspace{-0.3em }\ref{fig:aggregation} (c). This is due to a similar mechanism that occurs in the internal region; the high density making the whole process slower and resulting in a weaker aggregate. The noisiness of the data for the ABM makes it difficult to see such a weak aggregation. As time evolves, more agents in the internal region escape the two barriers (and some of the cells forming the barriers reorient) and reach the external region. The two barriers slowly move towards the centre and eventually coalesce. 

The aggregation phenomenon appears in both the PDE and the ABM (Fig. \hspace{-0.3em }\ref{fig:aggregation} (a)-(b)). However, the high level of spatial correlation associated with the aggregation, affects the quality of the agreement between the continuum and discrete models so the agreement is only qualitative and not quantitative. 

\citet{simpson2009mss} observed a form of spontaneous non-monotonicity in the continuous description of their multi-species on-lattice ABM. The authors considered a population of two identical, but distinctly labelled, species of cells moving according to a simple excluding random walk. The two species are initially confined in two adjacent regions with different initial densities. As time evolves, a non-monotonicity appears in the continuum profile of the species with lower density as consequence of the high density of the other species in the adjacent region. No clear evidence of such behaviour are present in the corresponding ABM. Moreover, it should be noted that any form of cell aggregation at the total population-level is not possible in the model of \citet{simpson2009mss}, since the overall behaviour is governed by simple diffusion. Although the spontaneous formation of aggregates (jamming) has been described previously in models which incorporate persistence  \citep{slowman2016jai,thompson2011lmn,sepulveda2016ccr,soto2014rtd,peruani2006ncs,levis2014chd,redner2013sdp,bialke2013mtp}, to the best of our our knowledge, non-monotonicity in average agent density has not been reported previously.

\section{Conclusion}
\label{sec:conclusion}
At the agent-level, we have modelled persistence of motion for interacting agents through excluding velocity-jump processes. Traditionally, such processes are associated with systems of advective PDEs which describes the model at a population scale. Our work continues the investigation of this type of model. First of all, we generalised the traditional velocity-jump model which allowed us to derive a system of diffusive equations from the ABMs (equations \eqref{eq:PDE_0}-\eqref{eq:PDE_1}-\eqref{eq:PDE_2} and \eqref{eq:PDE_3}). Moreover, our observations reveal some unusual phenomena (density spikes, anisotropy and aggregation) caused by the interplay of persistence and volume exclusion.

Despite the new diffusive PDEs correctly predicting the macroscopic behaviour of the ABM for a wide range of parameters, (including large values of reorienting rate) our findings highlight unrealistic behaviours for certain choices of parameters. Specifically, when the agent velocity is large and we implement one of the two more complex forms of agent interaction (types 2-3), the density profile of the ABM presents regular peaks (spikes) in density which are not captured by the corresponding continuum models. We also give evidence of an inherent anisotropy that occurs when we implement persistence in an on-lattice context in two or more dimensions \cite{thompson2011lmn}. This phenomenon represents a problem when applying the model to experimental data, for which isotropy is usually a natural feature. One possibility for reducing the scale of this issue and still obtaining the macroscopic description would be to work on an hexagonal lattice. This would increase the number of preferential directions from four to six, making the anisotropy less evident, although not completely removing it. Alternatively, one could allow cells to move in diagonal directions as in \cite{thompson2011lmn} which may also serve to mitigate, but not completely remove anisotropy. Off-lattice models should have the advantage that they are not afflicted by anisotropy. However, the derivation of a corresponding macroscopic description becomes more complicated and sometimes intractable. Therefore, the problem of modelling persistence of motion in an on-lattice context at multiple scales, without incurring anisotropy in the lattice directions, remains an interesting challenge for future research.

Finally, the other main achievement of this work is that the new continuum approximation that we propose is capable of qualitatively reproducing the spontaneous aggregation driven by persistence and volume exclusion \cite{slowman2016jai,thompson2011lmn,sepulveda2016ccr,soto2014rtd,peruani2006ncs,levis2014chd,redner2013sdp,bialke2013mtp}. To our knowledge, this is the first time that such behaviour in the ABM has been replicated at the macroscopic level.  
 The unintuitive consequence is that, in the case of strong persistence, the process of cell dispersion is initially slowed down by the aggregation phenomenon which constrains some of the cells in the internal region around the initial condition. Once the aggregates dissolve, the agents' dispersion is effectively faster than the normal diffusion. When such aggregation occurs, the agreement between the models at the two scales is qualitative rather than quantitative. More work might be done in order to recover a better agreement. In particular, this might be achieved by including a higher order of spatial correlations in the continuous model as in \citet{markham2013isc,markham2013smi}, however this remains an open question.\\

\ack
%This research was funded by 
The authors would like to thank the CMB/CNCB preprint club for constructive and helpful comments on a preprint of this paper.
%--------------------------------------------------------------------------------
% Bibliography 
\bibliography{database}{}
%\bibliographystyle{unsrtnat}
%\bibliographystyle{apsrev4-1}

%--------------------------------------------------------------------------------
\appendix \newpage
\appendixname
\renewcommand{\thesection}{\Alph{section}}
\numberwithin{equation}{section}

\section{Occupancy Master Equations}
\label{sec:OME}
In this section we report the occupancy master equations for the ABM described in Section \ref{sec:ABM} in the main text for the three types of non-trivial forms of agent interaction. 

\subsection*{Type 0} The full set of occupancy master equations for agents moving according to type 0 interactions are reported in the main document (see system \eqref{eq:master_0} in Section \ref{sec:cont_approx} of the main text).
\subsection*{Type 1}
For type 1 agent interactions, which corresponds to the case in which the movement is aborted if, and only if, the target site is occupied, the occupancy master equations read as follow:
\label{sec:supp_PDE_b}
\begin{equation}
\begin{split}
	\label{eq:master_1}
	%equation for R
	&\begin{aligned}
	R^{t+\tau}_{i,j}=R^t_{i,j}&+\frac{\tau P_m}{4} \left(1-C^t_{i,j}\right) \left[ (1+\varphi )R^t_{i-v,j}+ (1-\varphi)R^t_{i+v,j} +  R^t_{i,j+v}+R^t_{i,j-v} \right] \\
	& -\frac{\tau P_m}{4} R^t_{i,j} \left[ (1+\varphi ) \left(1-C^t_{i+v,j}\right)+ (1-\varphi)\left(1-C^t_{i-v,j}\right) +  \left(1-C^t_{i,j+v}\right) +\left(1-C^t_{i,j-v}\right)\right]\\
	&+ \frac{\tau P_r}{4} \left[ L^t_{i,j}+U^t_{i,j}+D^t_{i,j}-3R^t_{i,j}\right] +\mathcal{O}(\tau^2),
	\end{aligned} \\[15pt]
	%equation for L
	&\begin{aligned}
	L^{t+\tau}_{i,j}=L^t_{i,j}&+\frac{\tau P_m}{4} \left(1-C^t_{i,j}\right) \left[ (1-\varphi )L^t_{i-v,j}+ (1+\varphi)L^t_{i+v,j} +  L^t_{i,j+v}+L^t_{i,j-v} \right] \\
	& -\frac{\tau P_m}{4} L^t_{i,j} \left[ (1-\varphi ) \left(1-C^t_{i+v,j}\right)+ (1+\varphi)\left(1-C^t_{i-v,j}\right) +  \left(1-C^t_{i,j+v}\right) +\left(1-C^t_{i,j-v}\right)\right]\\
	&+ \frac{\tau P_r}{4} \left[ R^t_{i,j}+U^t_{i,j}+D^t_{i,j}-3L^t_{i,j}\right] +\mathcal{O}(\tau^2),
	\end{aligned}\\[15pt]
	%equation for U
	&\begin{aligned}
	U^{t+\tau}_{i,j}=U^t_{i,j}&+\frac{\tau P_m}{4} \left(1-C^t_{i,j}\right) \left[ (U^t_{i-v,j}+ U^t_{i+v,j} +  (1-\varphi) U^t_{i,j+v}+(1+\varphi) U^t_{i,j-v} \right] \\
	& -\frac{\tau P_m}{4} U^t_{i,j} \left[ \left(1-C^t_{i+v,j}\right)+ \left(1-C^t_{i-v,j}\right) +  (1+\varphi ) \left(1-C^t_{i,j+v}\right) + (1-\varphi )\left(1-C^t_{i,j-v}\right)\right]\\
	&+ \frac{\tau P_r}{4} \left[ R^t_{i,j}+L^t_{i,j}+D^t_{i,j}-3U^t_{i,j}\right] +\mathcal{O}(\tau^2),
	\end{aligned}\\[15pt]
	%equation for D
	&\begin{aligned}
	D^{t+\tau}_{i,j}=D^t_{i,j}&+\frac{\tau P_m}{4} \left(1-C^t_{i,j}\right) \left[ (D^t_{i-v,j}+ D^t_{i+v,j} +  (1+\varphi) D^t_{i,j+v}+(1-\varphi) D^t_{i,j-v} \right] \\
	& -\frac{\tau P_m}{4} D^t_{i,j} \left[ \left(1-C^t_{i+v,j}\right)+ \left(1-C^t_{i-v,j}\right) +  (1-\varphi ) \left(1-C^t_{i,j+v}\right) + (1+\varphi )\left(1-C^t_{i,j-v}\right)\right]\\
	&+ \frac{\tau P_r}{4} \left[ R^t_{i,j}+L^t_{i,j}+U^t_{i,j}-3D^t_{i,j}\right] +\mathcal{O}(\tau^2).
	\end{aligned}
	\end{split}
\end{equation}

\subsection*{Type 2}
\label{sec:supp_PDE_c}
Here we report the complete set of occupancy master equations for agents moving according to type 2 interactions. In this case agents move if, and only if, all the sites between the initial site and the target site are available. The equations are as follows:
\begin{equation}
\begin{split}
	\label{eq:master_2}
	%equation for R
	&\begin{aligned}
	R^{t+\tau}_{i,j}=R^t_{i,j}&+\frac{\tau P_m}{4}\Big[ (1+\varphi )R^t_{i-v,j} \prod_{s=0}^{v-1} \left(1-C^t_{i-s,j}\right) + (1-\varphi)R^t_{i+v,j} \prod_{s=0}^{v-1} \left(1-C^t_{i+s,j}\right)\\ 
	&+ R^t_{i,j+v}\prod_{s=0}^{v-1} \left(1-C^t_{i,j+s}\right)+R^t_{i,j-v}\prod_{s=0}^{v-1} \left(1-C^t_{i,j-s}\right) \Big] 
	 -\frac{\tau P_m}{4} R^t_{i,j} \Big[ (1+\varphi ) \prod_{s=1}^v \left(1-C^t_{i+s,j}\right)\\&+ (1-\varphi)\prod_{s=1}^v \left(1-C^t_{i-s,j}\right) +  \prod_{s=1}^v \left(1-C^t_{i,j+s}\right) +\prod_{s=1}^v \left(1-C^t_{i,j-s}\right)\Big]\\
	&+ \frac{\tau P_r}{4} \left[ L^t_{i,j}+U^t_{i,j}+D^t_{i,j}-3R^t_{i,j}\right] +\mathcal{O}(\tau^2),
	\end{aligned} \\[15pt] 	
	%equation for L
	&\begin{aligned}
	L^{t+\tau}_{i,j}=L^t_{i,j}&+\frac{\tau P_m}{4}\Big[ (1-\varphi )L^t_{i-v,j} \prod_{s=0}^{v-1} \left(1-C^t_{i-s,j}\right) + (1+\varphi)L^t_{i+v,j} \prod_{s=0}^{v-1} \left(1-C^t_{i+s,j}\right)\\ 
	&+ L^t_{i,j+v}\prod_{s=0}^{v-1} \left(1-C^t_{i,j+s}\right)+L^t_{i,j-v}\prod_{s=0}^{v-1} \left(1-C^t_{i,j-s}\right) \Big] 
	 -\frac{\tau P_m}{4} L^t_{i,j} \Big[ (1-\varphi ) \prod_{s=1}^v \left(1-C^t_{i+s,j}\right)\\&+ (1+\varphi)\prod_{s=1}^v \left(1-C^t_{i-s,j}\right) +  \prod_{s=1}^v \left(1-C^t_{i,j+s}\right) +\prod_{s=1}^v \left(1-C^t_{i,j-s}\right)\Big]\\
	&+ \frac{\tau P_r}{4} \left[ R^t_{i,j}+U^t_{i,j}+D^t_{i,j}-3L^t_{i,j}\right] +\mathcal{O}(\tau^2),	\end{aligned}\\[15pt]
	%equation for U
	&\begin{aligned}
	U^{t+\tau}_{i,j}=U^t_{i,j}&+\frac{\tau P_m}{4}\Big[ U^t_{i-v,j} \prod_{s=0}^{v-1} \left(1-C^t_{i-s,j}\right) + U^t_{i+v,j} \prod_{s=0}^{v-1} \left(1-C^t_{i+s,j}\right)
	+ (1-\varphi)U^t_{i,j+v}\prod_{s=0}^{v-1} \left(1-C^t_{i,j+s}\right)\\&+(1+\varphi) U^t_{i,j-v}\prod_{s=0}^{v-1} \left(1-C^t_{i,j-s}\right) \Big] 
	 -\frac{\tau P_m}{4} U^t_{i,j} \Big[  \prod_{s=1}^v \left(1-C^t_{i+s,j}\right)\\&+ \prod_{s=1}^v \left(1-C^t_{i-s,j}\right) +  (1+\varphi) \prod_{s=1}^v \left(1-C^t_{i,j+s}\right) +(1-\varphi) \prod_{s=1}^v \left(1-C^t_{i,j-s}\right)\Big]\\
	 	&+ \frac{\tau P_r}{4} \left[ R^t_{i,j}+L^t_{i,j}+D^t_{i,j}-3U^t_{i,j}\right] +\mathcal{O}(\tau^2),
	\end{aligned}\\[15pt]
	%equation for D
	&\begin{aligned}
	D^{t+\tau}_{i,j}=D^t_{i,j}&+\frac{\tau P_m}{4}\Big[ D^t_{i-v,j} \prod_{s=0}^{v-1} \left(1-C^t_{i-s,j}\right) + D^t_{i+v,j} \prod_{s=0}^{v-1} \left(1-C^t_{i+s,j}\right)
	+ (1+\varphi)D^t_{i,j+v}\prod_{s=0}^{v-1} \left(1-C^t_{i,j+s}\right)\\&+(1-\varphi) D^t_{i,j-v}\prod_{s=0}^{v-1} \left(1-C^t_{i,j-s}\right) \Big] 
	 -\frac{\tau P_m}{4} D^t_{i,j} \Big[  \prod_{s=1}^v \left(1-C^t_{i+s,j}\right)\\&+  \prod_{s=1}^v \left(1-C^t_{i-s,j}\right) +  (1-\varphi) \prod_{s=1}^v \left(1-C^t_{i,j+s}\right) +(1+\varphi)\prod_{s=1}^v \left(1-C^t_{i,j-s}\right)\Big]\\	&+ \frac{\tau P_r}{4} \left[ R^t_{i,j}+L^t_{i,j}+U^t_{i,j}-3D^t_{i,j}\right] +\mathcal{O}(\tau^2).
	\end{aligned}
	\end{split}
\end{equation}
\subsection*{Type 3}
\label{sec:supp_PDE_d}
Finally, we report the occupancy master equations for agents moving according to type 3 interactions, in which the agents move to the furthest available site. These read 
\begin{subequations}
\label{eq:master_3}
\begin{align}
		%equation for R
	&\begin{aligned}
	R^{t+\tau}_{i,j}=R^t_{i,j}&+\frac{\tau P_m}{4}\Big[ (1+\varphi )R^t_{i-v,j} \prod_{s=0}^{v-1} \left(1-C^t_{i-s,j}\right) + (1-\varphi)R^t_{i+v,j} \prod_{s=0}^{v-1} \left(1-C^t_{i+s,j}\right)\\ 
	&+ R^t_{i,j+v}\prod_{s=0}^{v-1} \left(1-C^t_{i,j+s}\right)+R^t_{i,j-v}\prod_{s=0}^{v-1} \left(1-C^t_{i,j-s}\right) \Big] 
	 -\frac{\tau P_m}{4} R^t_{i,j} \Big[ (1+\varphi ) \prod_{s=1}^v \left(1-C^t_{i+s,j}\right)\\&+ (1-\varphi)\prod_{s=1}^v \left(1-C^t_{i-s,j}\right) +  \prod_{s=1}^v \left(1-C^t_{i,j+s}\right) +\prod_{s=1}^v \left(1-C^t_{i,j-s}\right)\Big]\\
	& +\frac{\tau P_m}{4}\Big[ (1+\varphi )C^t_{i+1,j} \sum_{k=1}^{v-1} R^t_{i-k,j} \prod_{s=0}^{k-1} \left(1-C^t_{i-s,j}\right) + (1-\varphi)C^t_{i-1,j} \sum_{k=1}^{v-1} R^t_{i+k,j} \prod_{s=0}^{k-1} \left(1-C^t_{i+s,j}\right)\\ 
	&+ C^t_{i,j-1} \sum_{k=1}^{v-1} R^t_{i,j+k}\prod_{s=0}^{k-1} \left(1-C^t_{i,j+s}\right)+C^t_{i,j+1} \sum_{k=1}^{v-1} R^t_{i,j-k}\prod_{s=0}^{k-1} \left(1-C^t_{i,j-s}\right) \Big] \\ 
	&-\frac{\tau P_m}{4} R^t_{i,j} \Big[ (1+\varphi ) \sum_{k=2}^v \prod_{s=1}^{k-1} C^t_{i+k,j} \left(1-C^t_{i+s,j}\right)+ (1-\varphi)\sum_{k=1}^v \prod_{s=1}^{k-1} C^t_{i-k,j} \left(1-C^t_{i-s,j}\right) \\ 
	&+ \sum_{k=2}^v \prod_{s=1}^{k-1} C^t_{i,j+k} \left(1-C^t_{i,j+s}\right) +\sum_{k=2}^v\prod_{s=1}^{k-1} C^t_{i,j-k} \left(1-C^t_{i,j-s}\right)\Big]\\
	&+ \frac{\tau P_r}{4} \left[ L^t_{i,j}+U^t_{i,j}+D^t_{i,j}-3R^t_{i,j}\right] +\mathcal{O}(\tau^2),
	\end{aligned}\nonumber  \\[15pt] 	
	%equation for L
	&\begin{aligned}
	L^{t+\tau}_{i,j}=L^t_{i,j}&+\frac{\tau P_m}{4}\Big[ (1-\varphi )L^t_{i-v,j} \prod_{s=0}^{v-1} \left(1-C^t_{i-s,j}\right) + (1+\varphi)L^t_{i+v,j} \prod_{s=0}^{v-1} \left(1-C^t_{i+s,j}\right)\\ 
	&+ L^t_{i,j+v}\prod_{s=0}^{v-1} \left(1-C^t_{i,j+s}\right)+L^t_{i,j-v}\prod_{s=0}^{v-1} \left(1-C^t_{i,j-s}\right) \Big] 
	 -\frac{\tau P_m}{4} L^t_{i,j} \Big[ (1-\varphi ) \prod_{s=1}^v \left(1-C^t_{i+s,j}\right)\\&+ (1+\varphi)\prod_{s=1}^v \left(1-C^t_{i-s,j}\right) +  \prod_{s=1}^v \left(1-C^t_{i,j+s}\right) +\prod_{s=1}^v \left(1-C^t_{i,j-s}\right)\Big]\\
	& +\frac{\tau P_m}{4}\Big[ (1-\varphi )C^t_{i+1,j} \sum_{k=1}^{v-1} L^t_{i-k,j} \prod_{s=0}^{k-1} \left(1-C^t_{i-s,j}\right) + (1+\varphi)C^t_{i-1,j} \sum_{k=1}^{v-1} L^t_{i+k,j} \prod_{s=0}^{k-1} \left(1-C^t_{i+s,j}\right)\\ 
	&+ C^t_{i,j-1} \sum_{k=1}^{v-1} L^t_{i,j+k}\prod_{s=0}^{k-1} \left(1-C^t_{i,j+s}\right)+C^t_{i,j+1} \sum_{k=1}^{v-1} L^t_{i,j-k}\prod_{s=0}^{k-1} \left(1-C^t_{i,j-s}\right) \Big] \\ 
	&-\frac{\tau P_m}{4} L^t_{i,j} \Big[ (1-\varphi ) \sum_{k=2}^v \prod_{s=1}^{k-1} C^t_{i+k,j} \left(1-C^t_{i+s,j}\right)+ (1+\varphi)\sum_{k=1}^v \prod_{s=1}^{k-1} C^t_{i-k,j} \left(1-C^t_{i-s,j}\right) \\ 
	&+ \sum_{k=2}^v \prod_{s=1}^{k-1} C^t_{i,j+k} \left(1-C^t_{i,j+s}\right) +\sum_{k=2}^v\prod_{s=1}^{k-1} C^t_{i,j-k} \left(1-C^t_{i,j-s}\right)\Big]\\
	&+ \frac{\tau P_r}{4} \left[ R^t_{i,j}+U^t_{i,j}+D^t_{i,j}-3L^t_{i,j}\right] +\mathcal{O}(\tau^2),	\end{aligned}
	\displaybreak[4]\\[15pt]
	%equation for U
	&\begin{aligned}
	U^{t+\tau}_{i,j}=U^t_{i,j}&+\frac{\tau P_m}{4}\Big[ U^t_{i-v,j} \prod_{s=0}^{v-1} \left(1-C^t_{i-s,j}\right) + U^t_{i+v,j} \prod_{s=0}^{v-1} \left(1-C^t_{i+s,j}\right)
	+ (1-\varphi)U^t_{i,j+v}\prod_{s=0}^{v-1} \left(1-C^t_{i,j+s}\right)\\&+(1+\varphi) U^t_{i,j-v}\prod_{s=0}^{v-1} \left(1-C^t_{i,j-s}\right) \Big] 
	 -\frac{\tau P_m}{4} U^t_{i,j} \Big[  \prod_{s=1}^v \left(1-C^t_{i+s,j}\right)\\&+ \prod_{s=1}^v \left(1-C^t_{i-s,j}\right) +  (1+\varphi) \prod_{s=1}^v \left(1-C^t_{i,j+s}\right) +(1-\varphi) \prod_{s=1}^v \left(1-C^t_{i,j-s}\right)\Big]\\
	 & +\frac{\tau P_m}{4}\Big[ C^t_{i+1,j} \sum_{k=1}^{v-1} U^t_{i-k,j} \prod_{s=0}^{k-1} \left(1-C^t_{i-s,j}\right) + C^t_{i-1,j} \sum_{k=1}^{v-1} U^t_{i+k,j} \prod_{s=0}^{k-1} \left(1-C^t_{i+s,j}\right)\\ 
	&+ (1-\varphi )C^t_{i,j-1} \sum_{k=1}^{v-1} U^t_{i,j+k}\prod_{s=0}^{k-1} \left(1-C^t_{i,j+s}\right)+(1+\varphi )C^t_{i,j+1} \sum_{k=1}^{v-1} U^t_{i,j-k}\prod_{s=0}^{k-1} \left(1-C^t_{i,j-s}\right) \Big] \\ 
	&-\frac{\tau P_m}{4} U^t_{i,j} \Big[ \sum_{k=2}^v \prod_{s=1}^{k-1} C^t_{i+k,j} \left(1-C^t_{i+s,j}\right)+ \sum_{k=1}^v \prod_{s=1}^{k-1} C^t_{i-k,j} \left(1-C^t_{i-s,j}\right) \\ 
	&+ (1+\varphi ) \sum_{k=2}^v \prod_{s=1}^{k-1} C^t_{i,j+k} \left(1-C^t_{i,j+s}\right) +(1-\varphi ) \sum_{k=2}^v\prod_{s=1}^{k-1} C^t_{i,j-k} \left(1-C^t_{i,j-s}\right)\Big]\\
	 	&+ \frac{\tau P_r}{4} \left[ R^t_{i,j}+L^t_{i,j}+D^t_{i,j}-3U^t_{i,j}\right] +\mathcal{O}(\tau^2),
	\end{aligned}\nonumber\\[15pt]
	%equation for D
	&\begin{aligned}
	D^{t+\tau}_{i,j}=D^t_{i,j}&+\frac{\tau P_m}{4}\Big[ D^t_{i-v,j} \prod_{s=0}^{v-1} \left(1-C^t_{i-s,j}\right) + D^t_{i+v,j} \prod_{s=0}^{v-1} \left(1-C^t_{i+s,j}\right)
	+ (1+\varphi)D^t_{i,j+v}\prod_{s=0}^{v-1} \left(1-C^t_{i,j+s}\right)\\&+(1-\varphi) D^t_{i,j-v}\prod_{s=0}^{v-1} \left(1-C^t_{i,j-s}\right) \Big] 
	 -\frac{\tau P_m}{4} D^t_{i,j} \Big[  \prod_{s=1}^v \left(1-C^t_{i+s,j}\right)\\&+  \prod_{s=1}^v \left(1-C^t_{i-s,j}\right) +  (1-\varphi) \prod_{s=1}^v \left(1-C^t_{i,j+s}\right) +(1+\varphi)\prod_{s=1}^v \left(1-C^t_{i,j-s}\right)\Big]\\
	 & +\frac{\tau P_m}{4}\Big[ C^t_{i+1,j} \sum_{k=1}^{v-1} D^t_{i-k,j} \prod_{s=0}^{k-1} \left(1-C^t_{i-s,j}\right) + C^t_{i-1,j} \sum_{k=1}^{v-1} D^t_{i+k,j} \prod_{s=0}^{k-1} \left(1-C^t_{i+s,j}\right)\\ 
	&+ (1+\varphi )C^t_{i,j-1} \sum_{k=1}^{v-1} D^t_{i,j+k}\prod_{s=0}^{k-1} \left(1-C^t_{i,j+s}\right)+(1-\varphi )C^t_{i,j+1} \sum_{k=1}^{v-1} D^t_{i,j-k}\prod_{s=0}^{k-1} \left(1-C^t_{i,j-s}\right) \Big] \\ 
	&-\frac{\tau P_m}{4} D^t_{i,j} \Big[ \sum_{k=2}^v \prod_{s=1}^{k-1} C^t_{i+k,j} \left(1-C^t_{i+s,j}\right)+ \sum_{k=1}^v \prod_{s=1}^{k-1} C^t_{i-k,j} \left(1-C^t_{i-s,j}\right) \\ 
	&+ (1-\varphi ) \sum_{k=2}^v \prod_{s=1}^{k-1} C^t_{i,j+k} \left(1-C^t_{i,j+s}\right) +(1+\varphi ) \sum_{k=2}^v\prod_{s=1}^{k-1} C^t_{i,j-k} \left(1-C^t_{i,j-s}\right)\Big]\\
	&+ \frac{\tau P_r}{4} \left[ R^t_{i,j}+L^t_{i,j}+U^t_{i,j}-3D^t_{i,j}\right] +\mathcal{O}(\tau^2).
	\end{aligned}
\end{align}	
\end{subequations}

\section{Complete Density Dystems}
\label{sec:CDS}

In this section we report the complete systems of PDEs for the four types of agent interaction. 

\subsection*{Type 0} The full set of PDEs for model with type 0 interactions is reported in the main document (see system \eqref{eq:PDE_0} in Section \ref{sec:cont_approx} of the main text).

\subsection*{Type 1}
For the type 1 form of interaction the system of diffusive PDEs for the four subpopulation reads as follows:
\begin{equation}
\begin{split}
	\label{eq:PDE_1}
	%equation for R
	&\begin{aligned}
	\D{R}{t}=v^2\mu \left[R \nabla^2 C+(1-C) \nabla^2 R\right] - v\nu \D{}{x} \left[ R(1-C)\right]+\frac{P_r}{4}(C -4 R) 
	,\end{aligned} \\
	%equation for L
	&\begin{aligned}
	\D{L}{t}=v^2\mu \left[L \nabla^2 C+(1-C) \nabla^2 L\right] + v\nu \D{}{x} \left[ L(1-C)\right]+\frac{P_r}{4}(C -4 L) 
	,\end{aligned} \\
	%equation for U
	&\begin{aligned}
	\D{U}{t}=v^2\mu \left[U \nabla^2 C+(1-C) \nabla^2 U\right] - v\nu \D{}{y} \left[ U(1-C)\right]+\frac{P_r}{4}(C -4 U) 
	,\end{aligned} \\
	%equation for D
	&\begin{aligned}
	\D{D}{t}=v^2\mu \left[D \nabla^2 C+(1-C) \nabla^2 D\right] + v\nu \D{}{y} \left[ D(1-C)\right]+\frac{P_r}{4}(C -4 D) 
	,\end{aligned} \\
	\end{split}
\end{equation}
where $\mu$ and $\nu$ are defined as in equations \eqref{eq:coeff} and the boundary conditions are imposed as in equations \eqref{eq:BC_PDE} of the main text.
\subsection*{Type 2}
For the type 2 interaction, the complete set of PDEs is
\begin{equation}
\begin{split}
	\label{eq:PDE_2}
	%equation for R
	&\begin{aligned}
	\D{R}{t}=&v^2\mu \left[R \nabla \left( (1-C)^{v-1} \nabla C\right)+(1-C)\nabla \left( (1-C)^{v-1} \nabla R\right)\right]\\& - v\nu \D{}{x} \left[ R(1-C)^v\right]+\frac{P_r}{4}(C -4 R) 
	,\end{aligned} \\
	%equation for L
	&\begin{aligned}
	\D{L}{t}=&v^2\mu \left[L \nabla \left( (1-C)^{v-1} \nabla C\right)+(1-C)\nabla \left( (1-C)^{v-1} \nabla L\right)\right]\\ &  + v\nu \D{}{x} \left[ L(1-C)^v\right]+\frac{P_r}{4}(C -4 L) 
	,\end{aligned} \\
	%equation for U
	&\begin{aligned}
	\D{U}{t}=&v^2\mu \left[U \nabla \left( (1-C)^{v-1} \nabla C\right)+(1-C)\nabla \left( (1-C)^{v-1} \nabla U\right)\right]\\ &  - v\nu \D{}{y} \left[ U(1-C)^v\right]+\frac{P_r}{4}(C -4 U) 
	,\end{aligned} \\
	%equation for D
	&\begin{aligned}
	\D{D}{t}=&v^2\mu \left[D \nabla \left( (1-C)^{v-1} \nabla C\right)+(1-C)\nabla \left( (1-C)^{v-1} \nabla D\right)\right]\\ &  +v \nu \D{}{y} \left[ D(1-C)^v\right]+\frac{P_r}{4}(C -4 D) 
	,\end{aligned} \\
	\end{split}
\end{equation}
where $\mu$ and $\nu$ are defined as in equations \eqref{eq:coeff} and the boundary conditions are imposed as in equations \eqref{eq:BC_PDE} of the main text.
\subsection*{Type 3}

Finally, we report the complete diffusive system of for the model with type 3 interaction. This reads
\begin{equation}
\begin{split}
	\label{eq:PDE_3}
	%equation for R
	&\begin{aligned}
	\D{R}{t}=&\mu \nabla \left[\sum_{k=1}^v (1-C)^{k-1}  \left[ (2k-1)(1-C) \nabla R -k (k-2) R \nabla C \right] \right] \\& + \nu \D{}{x} \left[ \frac{(1-C)\left( (1-C)^v -1\right) }{C} R\right] +\frac{P_r}{4}(C -4 R) 
	,\end{aligned} \\
	%equation for L
	&\begin{aligned}
	\D{L}{t}=&\mu \nabla \left[\sum_{k=1}^v (1-C)^{k-1}  \left[ (2k-1)(1-C) \nabla L -k (k-2) L \nabla C \right] \right] \\& - \nu \D{}{x} \left[ \frac{(1-C)\left( (1-C)^v -1\right) }{C} L\right] +\frac{P_r}{4}(C -4 L) 
	,\end{aligned} \\
	%equation for U
	&\begin{aligned}
	\D{U}{t}=&\mu \nabla \left[\sum_{k=1}^v (1-C)^{k-1}  \left[ (2k-1)(1-C) \nabla U -k (k-2) U \nabla C \right] \right] \\& + \nu \D{}{y} \left[ \frac{(1-C)\left( (1-C)^v -1\right) }{C} U\right] +\frac{P_r}{4}(C -4 U) 
		,\end{aligned} \\
	%equation for D
	&\begin{aligned}
	\D{D}{t}=&\mu \nabla \left[\sum_{k=1}^v (1-C)^{k-1}  \left[ (2k-1)(1-C) \nabla D -k (k-2) D \nabla C \right] \right] \\& - \nu \D{}{y} \left[ \frac{(1-C)\left( (1-C)^v -1\right) }{C} D\right] +\frac{P_r}{4}(C -4 D) 
		,\end{aligned} \\
	\end{split}
\end{equation}
where $\mu$ and $\nu$ are defined as in equations \eqref{eq:coeff} and the boundary conditions are imposed as in equations \eqref{eq:BC_PDE} of the main text.

\end{document}